\documentclass[aps,prd,reprint]{revtex4-1}

\usepackage{amssymb,amsmath,epsf}
\usepackage{graphicx}
\usepackage{dcolumn}
\usepackage{mathtools}
\usepackage{amsfonts}
\usepackage{wrapfig}

\usepackage{esint}
\usepackage[T1]{fontenc}
\usepackage[utf8]{inputenc}
\usepackage{listings}
\usepackage{color}
\usepackage{esint}
\usepackage[english]{babel}
\usepackage[T1]{fontenc}
\usepackage[utf8]{inputenc}

\usepackage{booktabs}

\begin{document} 

\title{Electron-Positron Cascade in Magnetospheres of Spinning Black Holes}
\author{Alexander L. Ford}
\email{alexlford@ku.edu}
\affiliation{Department of Physics and Astronomy, University of Kansas, Lawrence, KS 66045}

\author{Brett D. Keenan}
\altaffiliation{Currently at Los Alamos National Lab, Los Alamos, NM 87545}
\affiliation{Department of Physics and Astronomy, University of Kansas, Lawrence, KS 66045}
%\altaffiliation{Los Alamos National Lab, Los Alamos, NM 87544}
%\altaffiliation{Research done at the Department of Physics and Astronomy, University of Kansas, Lawrence, KS 66045}

\author{Mikhail V. Medvedev} 
\altaffiliation{On sabbatical leave from the Department of Physics and Astronomy, University of Kansas, Lawrence, KS 66045}
\affiliation{Institute for Theory and Computation, Harvard University, Cambridge, MA 02138}

%\affiliation{Institute for Theory and Computation, Harvard University, 60 Garden St., Cambridge, MA 02138}
%\altaffiliation{Also at the Department of Physics and Astronomy, University of Kansas, Lawrence, KS 66045}

\date{\today}

\begin{abstract}

We quantitatively study the stationary, axisymmetric, force-free magnetospheres of spinning (Kerr) black holes (BHs) and the conditions needed for relativistic jets to be powered by the Blandford-Znajek mechanism. These jets could be from active galactic nuclei, blazars, quasars, micro-quasars, radio active galaxies, and other systems that host Kerr BHs. The structure of the magnetosphere determines how the BH energy is extracted, e.g., via Blandford-Znajek mechanism, which converts the BH rotational energy into Poynting flux. The key assumption is the force-free condition, which requires the presence of plasma with the density being above the Goldreich-Julian density. Unlike neutron stars, which in principle can supply electrons from the surface, BH cannot supply plasma at all. The plasma must be generated \text{\it in situ} via an electron-positron cascade, presumably in the gap region. Here we study varying conditions that provide a sufficient amount of plasma for the Blandford-Znajek mechanism to work effectively.
\\

\end{abstract}

\maketitle

\section{Introduction}

Since Blandford and Znajek's seminal paper, Ref. \cite{1977MNRAS.179..433B}, the plasma-rich magnetosphere around a Kerr black hole (BH) has been employed to explain how energy is extracted and used for powering jets. In the inner jet region, close to the BH, the magnetosphere must be force-free, i.e., $\rho_e\bf{E}+\bf{j}$$/c\times \bf{B}=0$, where $\rho_e$, $\bf{E}$, $\bf{j}$, $c$, and $\bf{B}$ are the charge density, electric field, current density, speed of light, and magnetic field, respectively. The mechanism for filling the magnetosphere with plasma has been discussed in previous works \cite{Beskin:1992va, Hirotani:1998cf, Levinson:2124471, Globus:1604932}. 

We assume a stationary, axisymmetric, force-free magnetosphere around a Kerr BH with mass $M$ and angular momentum $J$. We use Boyer-Lindquist coordinates ($t$, $r$, $\theta$, $\phi$) with the two scalar functions $\alpha$ and $\omega$ \cite{Thorne:1986iy}:
\begin{equation}
\begin{multlined}
ds^2=\left(\varpi^2\omega^2-\alpha^2\right)dt^2-2\omega\varpi^2d\phi dt+\\
\frac{\rho^2}{\Delta} dr^2+\rho^2d\theta^2+\varpi^2d\phi^2,
\end{multlined}
\end{equation}
where
\begin{equation}
\rho^2=r^2+a^2\cos^2\theta,
\end{equation}
\begin{equation}
\Delta=r^2+a^2-2Mr/c^2,
\end{equation}
\begin{equation}
\Sigma^2=(r^2+a^2)^2-a^2\Delta\sin^2\theta,
\end{equation}
\begin{equation}
\varpi=\frac{\Sigma}{\rho}\sin\theta,
\end{equation}
\begin{equation}
\alpha=\frac{\rho}{\Sigma}\sqrt{\Delta},
\end{equation}
\begin{equation}
\omega=\frac{2aGMr}{c\Sigma^2}.
\end{equation}
Here $a$ is the spin parameter of the BH, $a\equiv J/Mc$ and the BH radius is 
\begin{equation}
r_H = GM/c^2+\left[\left(GM/c^2\right)^2-a^2\right]^{1/2}.
\label{rH}	
\end{equation}
The redshift factor or the lapse function is $\alpha$ and $\omega$ is the angular velocity of the zero angular momentum observers (ZAMO), which coincides with uniform rotation of the BH and vanishes at infinity. In order to describe the electromagnetic processes, we use the 3+1 split of the laws of electrodynamics \cite{Thorne:1986iy}. We split the four-dimensional spacetime into a global time $t$ and an absolute three-dimensional curved space. 

The poloidal magnetic field may be expressed in terms of the magnetic flux function, $\Psi$ as follows \cite{Macdonald:1984cd},
\begin{equation}
\mathbf{B_p}=\frac{\nabla\Psi\times{\hat{\phi}}}{	2\pi\varpi}.
\end{equation}
Then using the force-free condition, $\mathbf{E}\cdot\mathbf{B}=0$ to find the poloidal electric field,
\begin{equation}
\mathbf{E_p}=\frac{\Omega_{F}-\omega}{2\pi\alpha c}\nabla\Psi	,
\end{equation}
where the velocity of the magnetic field lines from the ZAMO's reference frame is
\begin{equation}
\mathbf{v_F}=\frac{(\Omega_F-\omega)\varpi}{\alpha}{\hat{\phi}}	.
\end{equation}

\begin{figure}
\includegraphics[angle = 0, width = 0.9\columnwidth]{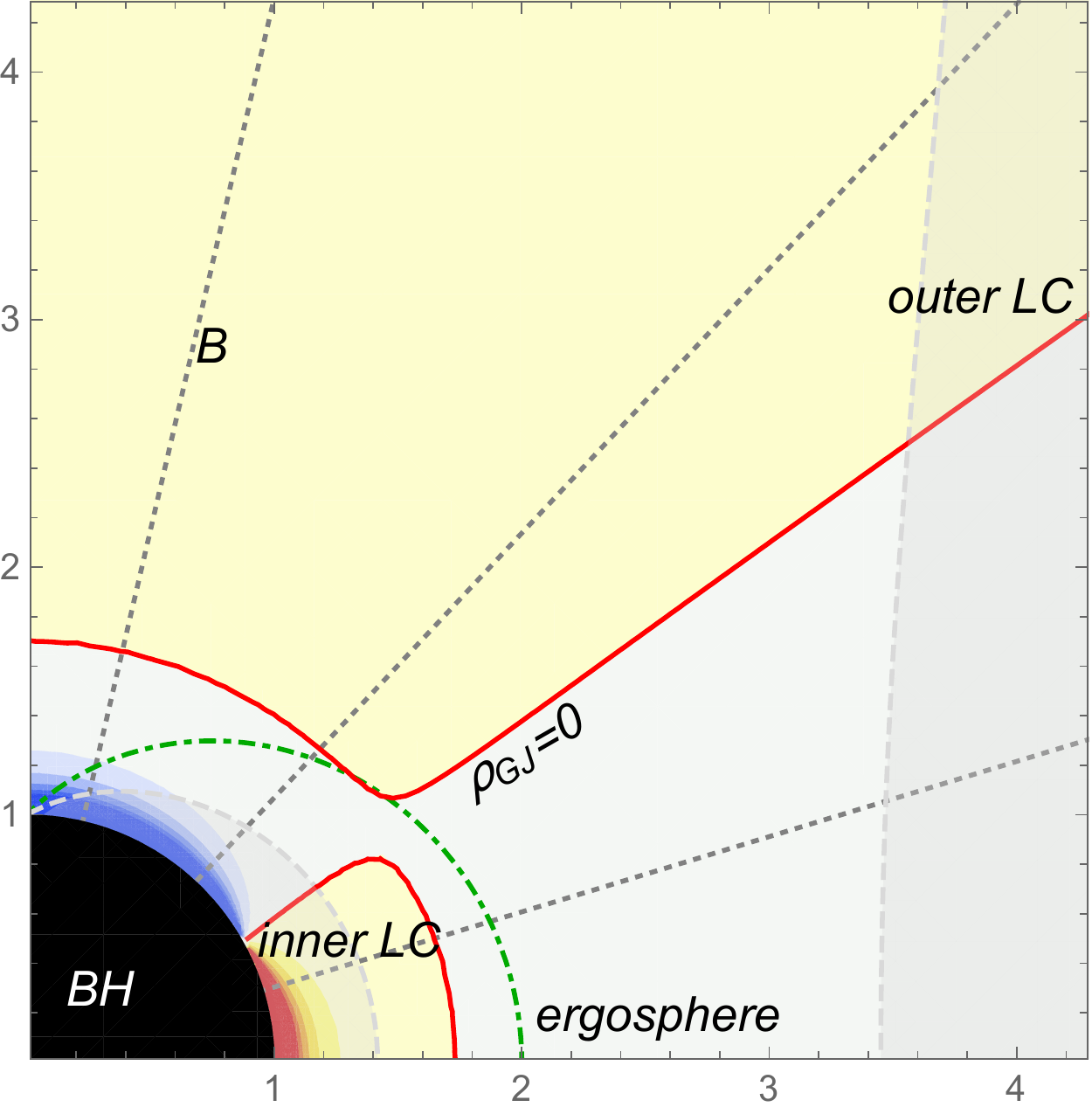}
%\vskip-0.3cm
\caption{The BH radius is set to one. The blue/gray regions and the red/yellow regions signify the plasma densities. The red, solid line is the surface where $\rho_{GJ}$ goes to zero. The green, dash-dotted line is the ergoshphere. The light gray, long-dashed lines represent the inner and outer light cylinder. And finally, the dark gray, short-dashed lines display the geometry of the magnetic field lines.}
\label{detail}
\end{figure}

The charge density needed for the degenerate magnetosphere to be force-free is:
\begin{equation}
\rho_{GJ}=\frac{1}{4\pi}\nabla\cdot \mathbf{E_p}=\frac{-1}{4\pi}\nabla\cdot\left(\frac{\Omega_F-\omega}{2\pi\alpha c}\nabla\Psi\right).
\end{equation}
Hereafter we assume the split monopole magnetic field configuration used in Ref. \cite{1977MNRAS.179..433B} with \begin{equation}
\Psi(\theta)= \Psi_M (1-cos(\theta)).
\end{equation}
There exists a surface where $\rho_{GJ}=0$, see Fig. \ref{detail}. In a force-free magnetosphere, this ``null surface''  has the potential to create a region with a strong electric field ($E_\parallel$) that is parallel to the magnetic field. The charge deficit around the ``null surface'' allows $E_\parallel$ to emerge, we will refer to this region simply as the gap.
%
%\begin{wrapfigure}{r}{0.601\textwidth}
%\vskip-0.5cm
%\vskip-0.3cm
%\end{wrapfigure}
%
Inside of the gap, the Poisson equation is:
\begin{equation}
\nabla\cdot E_\parallel=4\pi\left(\rho_e-\rho_{GJ}(x,\theta)\right),	
\end{equation}
where the charge density, 
$\rho_e\equiv e(n^+-n^-)$, is viewed in the corotating frame of the magnetic field and is the difference between positive ($n^+$) and negative ($n^-$) charges. As originally suggested in Ref. \cite{1977MNRAS.179..433B}, an electron-positron cascade is needed to maintain a force-free magnetosphere. Charged particles are accelerated by $E_\parallel$ inside of the gap, these accelerated particles inverse Compton scatter with background photons from, e.g., the accretion disk. This produces $\gamma$-rays which then collide with background photons and produce electron-positron pairs. These pairs in turn get accelerated and independently repeat the process until the magnetosphere is filled.

We extend previous works Refs. \cite{Beskin:1992va, Hirotani:1998cf} by looking at broad ranges of mass, magnetic field, background photon energy density, and spin. 

\section{Electron-Positron Cascade Mechanism}
In this section, we lay out the theoretical framework that governs how an electron-positron cascade occurs in the gap.
\subsection{Cascade Equations}
In the gap, there is insufficient plasma to screen out an electric field, that is why $E_\parallel$ emerges. We reduce the geometry to one dimension and rewrite the Poisson equation,
\begin{equation}
\frac{dE_\parallel}{dx}=4\pi\left[e\left(n^+-n^-\right)-\rho_{GJ}\right],	
\label{dE_org}
\end{equation}
where $x$ is perpendicular to the ``null surface'' and zero at the center of the gap, i.e., $x=(r-r_0)$, with $r_0$ being the ``null surface''. As will be shown, the gap is usually considerably smaller than $r_H$; therefore, we can expand $\rho_{GJ}(x,\theta)$ around $x=0$, the center of the gap. This allows us to rewrite the Poisson equation once again,
\begin{equation}
\frac{dE_\parallel}{dx}=4\pi\left[e\left(n^+-n^-\right)-A_\theta x\right],	
\label{dE}
\end{equation}
where $A_\theta$ is the expansion coefficient at a particular $\theta$, $A_\theta=\partial_r(\rho_{GJ}(x,\theta))$ at $x=0$.

Inside of the gap, the electrons/positrons ($e^\pm$s) will be accelerated by the $E_\parallel$ field. The motion of a single charge can be determined by:
\begin{equation}
m_ec^2\frac{d\Gamma}{dx}=eE_\parallel-\left(\Gamma^2-1\right)\sigma_TU_b	 ,
\label{dgamma}
\end{equation}
where $\Gamma$, $\sigma_T$, and $U_b$ are the Lorentz factor of the $e^\pm$, the Thomson cross section, and the energy density of the background photon field, respectively. These $e^\pm$s can produce $\gamma$-ray photons vis inverse Compton scattering with background photons \cite{Beskin:1992va}. The newly created $\gamma$-rays can now pair produce by colliding with other background photons. If a $\gamma$-ray with an energy $m_ec^2\epsilon_\gamma$ collides with a background photon with an energy $m_ec^2\epsilon_s$, then to produce an $e^\pm$ pair the energies must statisfy:
\begin{equation}
\epsilon_\gamma\epsilon_s\geq2/(1-\mu),	
\end{equation}
where $\mu$ is the cosine of the angle between the colliding photons.

Now considering the continuity equations for $e^\pm$, the direction of motion of the charges is set by the direction of the current, which is toward the BH in polar regions. The continuity equations are:

\begin{equation}
\begin{multlined}
\pm\frac{d}{dx}\left[n^\pm(x)\sqrt{1-\frac{1}{\Gamma^2(x)}}\right]=\\
\int_0^\infty \eta_p(\epsilon_\gamma)\left[F^+(x,\epsilon_\gamma)+F^-(x,\epsilon_\gamma)\right]d\epsilon_\gamma,
\end{multlined}
\label{dn}
\end{equation}
where $\eta_p$ is the angle-averaged pair production redistribution function and $F^\pm$ are the number densities of the $\gamma$-rays traveling in the $\pm x$ direction. 
At the boundary of the gap, $E_\parallel$ must go to zero. This only happens when $j_0=j_\text{critical}$, where $j_0$ is defined by:
\begin{equation}
j_0=e\left[n^+(x)+n^-(x)\right]\sqrt{1-1/\Gamma^2(x)}.
\label{j0}
\end{equation}
The critical current density is the constant outflow from the gap.
The $\gamma$-ray distribution functions, $F^\pm$, obey:
\begin{equation}
\begin{multlined}
\pm\frac{\partial}{\partial x}F^\pm(x,\epsilon_\gamma)=\\
\eta_c(\epsilon_\gamma,\Gamma(x))n^\pm(x)\sqrt{1+\frac{1}{\Gamma^2(x)}}-\eta_p(\epsilon_\gamma)F^\pm(x,\epsilon_\gamma),
\end{multlined}
\label{dF}
\end{equation}
where $\eta_c$ is the Compton redistribution function.
In order to numerically solve for the $\gamma$-ray distribution, $\epsilon_\gamma$ needs to be divided into energy bins. Let $\xi_i$ and $\xi_{i-1}$ be the upper and lower limits of the $i^\text{th}$ normalized energy bin. This allows us to rewrite the integral in Eq. \ref{dn} as 
\begin{equation}
\int_{\xi_{i-1}}^{\xi_i}\eta_p(\epsilon_\gamma)F^\pm(x,\epsilon_\gamma)d\epsilon_\gamma.
\end{equation}
Defining,
\begin{equation}
\eta_{p,i}\equiv\eta_p(\frac{\xi_i+\xi_{i-1}}{2}),
\label{etapi}	
\end{equation}
and
\begin{equation}
f_i^\pm(x)\equiv\int_{\xi_{i-1}}^{\xi_i}F^\pm(x,\epsilon_\gamma)d\epsilon_\gamma,
\label{fi}
\end{equation}
updating Eq. \ref{dn},
\begin{multline}
\pm\frac{d}{dx}\left\{n^\pm(x)\sqrt{1-\frac{1}{\Gamma^2(x)}}\right\}=\\
\sum^\chi_{i=1}\eta_{p,i}\left[f_i^+(x)+f_i^-(x)\right],
\label{npm}
\end{multline}
where $\chi$ is the number of normalized energy bins. An analogous approximation to Eq. \ref{fi} is implemented for $\eta_c$,
\begin{equation}
\eta_{c,i}\left(\Gamma(x)\right)\equiv\int_{\xi_{i-1}}^{\xi_i}\eta_c\left(\epsilon_\gamma,\Gamma(x)\right)d\epsilon_\gamma	,
\end{equation}
and allows us to express Eq. \ref{dF} as
\begin{multline}
\pm\frac{d}{dx}f_i^\pm(x)=\\
\eta_{c,i}(\Gamma(x))n^\pm(x)\sqrt{1+\frac{1}{\Gamma^2(x)}}-\eta_{p,i}f^\pm(x).
\label{fpm}	
\end{multline}

For all presented solutions, a power law spectrum for the background photon number density with an index of two is used. The dependence on the spectral index has been explored elsewhere \cite{Hirotani:1998cf}. The minimum and maximum energies of the background photon spectrum are 4.1 eV and 102 keV, respectively. After splitting $\epsilon_\gamma$ into $\chi$ discrete energy bins, we are left with 2$\chi$+3 ordinary differential equations (ODEs). Solving the ODEs with appropriate boundary conditions allows us to examine the structure of the gap. 

\subsection{Boundary Conditions}
The assumptions of symmetry that are used are as follows:
\begin{equation}
\begin{gathered}
E_\parallel(x)=E_\parallel(-x),\\
\Gamma(x)=\Gamma(-x),\\
n^+(x)=n^-(-x),\\
F^+(x)=F^-(-x).
\end{gathered}
\label{sym}
\end{equation}
These assumptions are applicable so long as the gap width stays small, $<1\%$ of the BH radius. Using these symmetries allows us to set the boundary conditions at the center of the gap and the edge of the gap; allowing us to only integrate over half of the gap and obtain a full solution. Using Eq. \ref{dgamma} with $E_\parallel(x)=E_\parallel(-x)$ and $\Gamma(x)=\Gamma(-x)$ at $x=0$ we get a boundary condition on $E_\parallel$
\begin{equation}
E_\parallel=\frac{\sigma_T U_b}{e}(\Gamma^2-1)	.
\end{equation}
Using Eq. \ref{j0} with $n^+(x)=n^-(-x)$ at $x=0$ yields another boundary condition
\begin{equation}
2n^+\sqrt{1-\frac{1}{\Gamma^2}}=\frac{j_0}{e}	.
\end{equation}
Using $F^+(x)=F^-(-x)$ at $x=0$ gives another boundary condition
\begin{equation}
f_i^+=f_i^- .
\end{equation}
The boundary of the gap is defined as the position when the plasma density in the gap is equal to $\rho_{GJ}$. Using Eq. \ref{dE_org} at $x=H$ we get a boundary condition on $E_\parallel$
\begin{equation}
E_\parallel=0 .
\end{equation}
$E_\parallel$ should go to zero smoothly at the boundary; therefore, $dE_\parallel/dx=0$ at $x=H$. Using this condition and Eq. \ref{j0} at   $x=H$ provides another boundary condition
\begin{equation}
j_0\left(1-\frac{1}{\Gamma^2}\right)^{-1/2}-A_\theta x=0 .
\end{equation}
Assuming that all of the charged particle are created inside of the gap; therefore, no charges should enter into the gap. Using $n^-=0$ and Eq. \ref{j0} at $x=H$ we get another boundary condition 
\begin{equation}
n^+\sqrt{1-\frac{1}{\Gamma^2}}=\frac{j_0}{e}	 .
\end{equation}
All upscatered photons are created inside of the gap. Assuming none will be coming into the gap we get another boundary condition
\begin{equation}
f_i^-=0 .
\end{equation}
This provides 2$\chi$+5 boundary conditions for 2$\chi$+3 ODEs and 2 constants: $j_0$ and $H$. These boundary conditions have be summarized in Table \ref{bc} for reference and clarity.
\begin{table*}
\renewcommand{\arraystretch}{2}
\begin{tabular}{|c|c|c|c|}
 \toprule
\hline
    Boundary Condition & Equation Used & Assumptions & Boundary\\
     \hline\hline
   
   $E_\parallel=(\Gamma^2-1)\sigma_T U_b/e$ & $m_ec^2d\Gamma/dx=eE_\parallel-\left(\Gamma^2-1\right)\sigma_TU_b$ & $E_\parallel(x)=E_\parallel(-x)$ \& $\Gamma(x)=\Gamma(-x)$ & $x=0$\\
     \hline
   $2n^+\sqrt{1-1/\Gamma^2}=j_0/e$ & $j_0=e\left[n^+(x)+n^-(x)\right]\sqrt{1-1/\Gamma^2(x)}$ & $n^+(x)=n^-(-x)$& $x=0$\\
     \hline
    $f_i^+=f_i^-$ & $f_i^\pm(x)\equiv\int_{\xi_{i-1}}^{\xi_i}F^\pm(x,\epsilon_\gamma)d\epsilon_\gamma$ & $F^+(x)=F^-(-x)$& $x=0$\\
     \hline
     $E_\parallel=0 $& $dE_\parallel/dx=4\pi\left[e\left(n^+-n^-\right)-\rho_{GJ}\right]$ & $\rho_{\text{gap}}=\rho_{GJ}$ & $x=H$\\
     \hline
     $n+\sqrt{1-1/\Gamma^2}=j_0/e$ & $j_0=e\left[n^+(x)+n^-(x)\right]\sqrt{1-1/\Gamma^2(x)}$ & $n^-(x)=0 $ & $x=H$\\
     \hline
     $j_0\left(1-1/\Gamma^2\right)^{-1/2}-Ax=0$ & $j_0=e\left[n^+(x)+n^-(x)\right]\sqrt{1-1/\Gamma^2(x)}$ & $dE_\parallel/dx=0 $ & $x=H$\\
     \hline
     $f_i^-=0$ & $f_i^-(x)\equiv\int_{\xi_{i-1}}^{\xi_i}F^-(x,\epsilon_\gamma)d\epsilon_\gamma$ & $F^-(x)=0$ & $x=H$\\
     \hline
      \bottomrule
\end{tabular}
\renewcommand{\arraystretch}{1}	
\caption{A complete overview of the boundary conditions and assumptions used to arrive at them.}
\label{bc}
\end{table*}

\section{Structure of the Gap}

\begin{figure}[!h]
\includegraphics[angle = 0, width = 0.9\columnwidth]{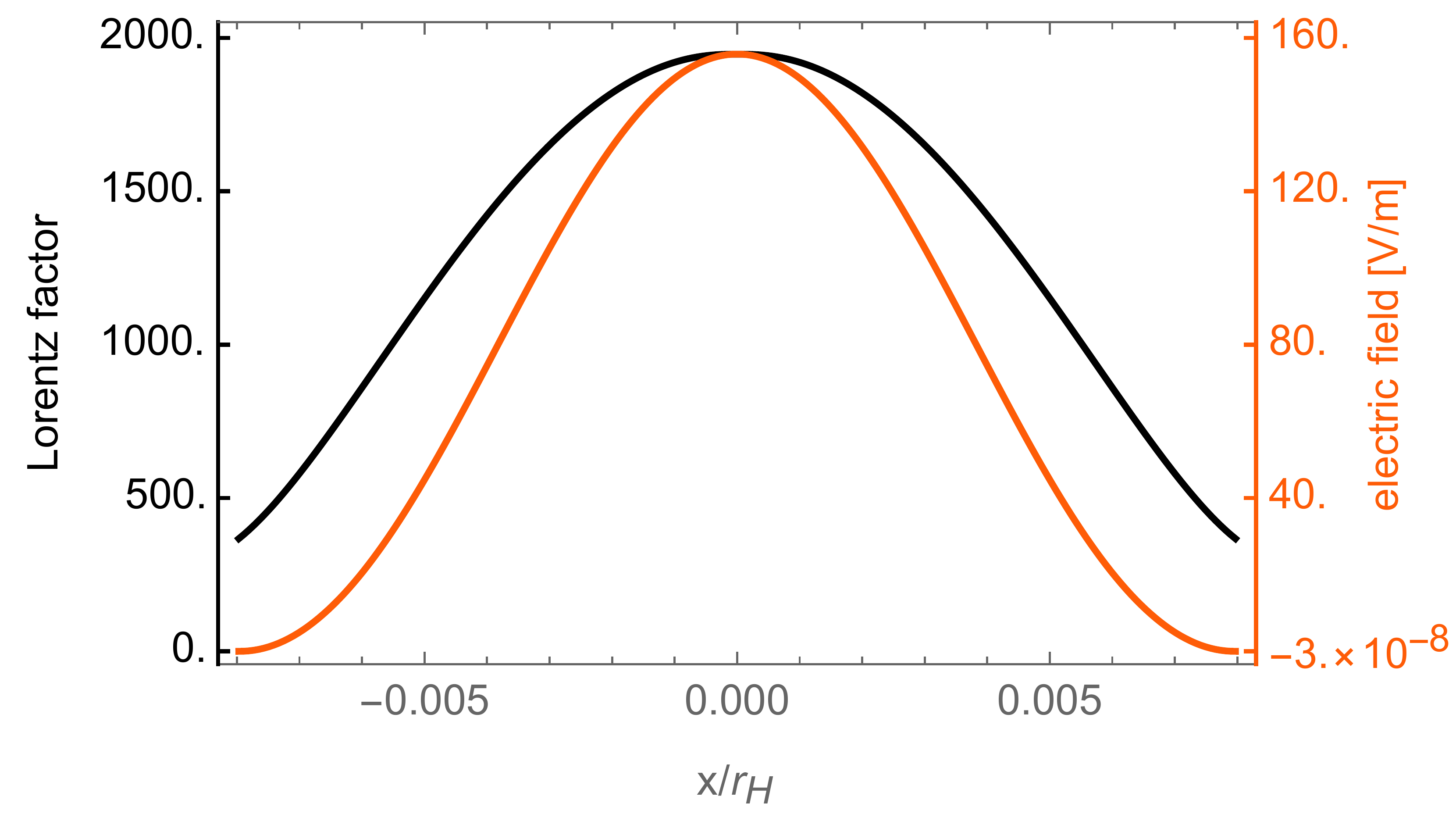}
\caption{The electric field in the gap and the Lorentz factor of the charges in the gap as a function of position inside of the gap. The $x$-axis has been normalized by the BH radius. This result is for a maximumly spinning BH of mass $10^7 M_\odot$ with a magnetic field strength of $10^4$ Gauss and an ambient energy density of $10^6 \text{ ergs}/\text{cm}^3$. The top curve shown in black represents the Lorentz factor and the orange curve represents the electric field.}
\label{1d_GandE}
\end{figure}

\begin{figure}[!h]
\includegraphics[angle = 0, width = 0.9\columnwidth]{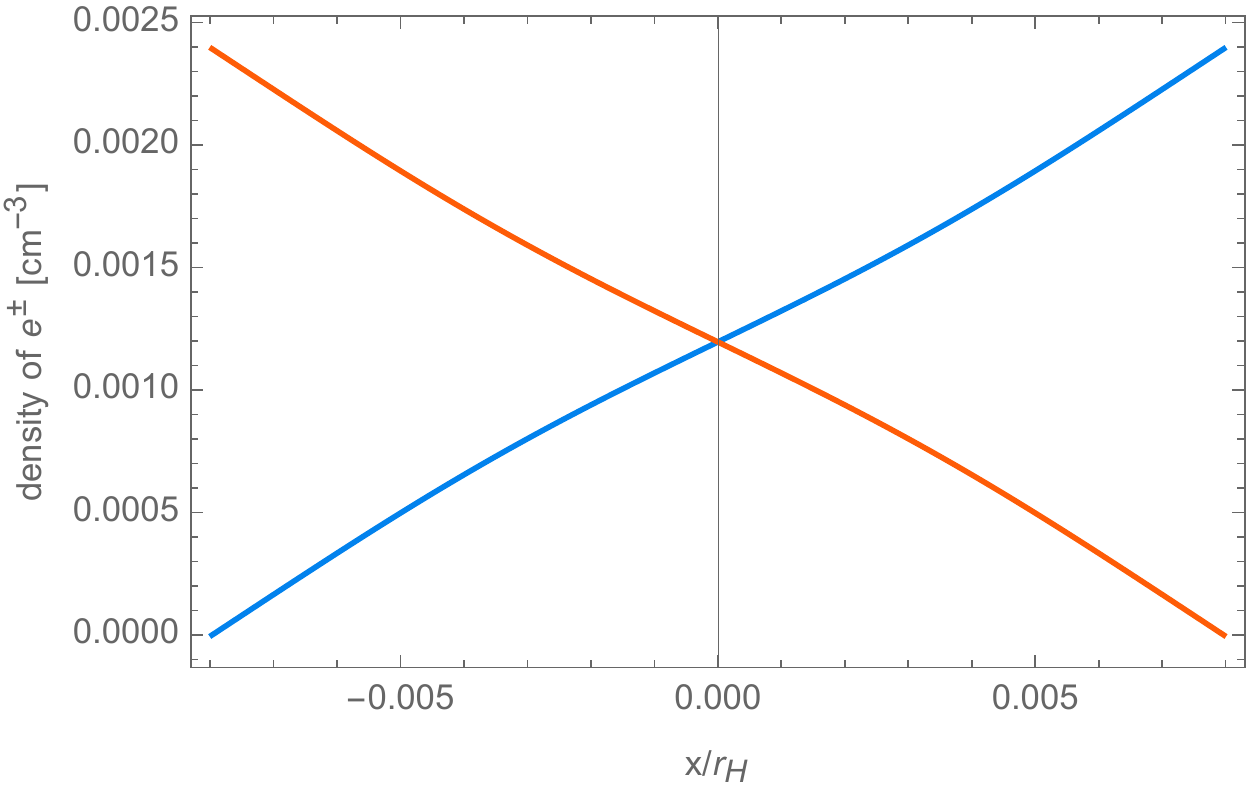}
\caption{The charge density as a function of position inside of the gap. The blue curve represents the charge density that is moving away from the BH and similarly the orange curve represents the inward moving charge density. The $x$-axis has been normalized by the BH radius. This result is for a maximumly spinning BH of mass $10^7 M_\odot$ with a magnetic field strength of $10^4$ Gauss and an ambient energy density of $10^6 \text{ ergs}/\text{cm}^3$.}
\label{1d_pairs}
\end{figure}

\begin{figure}[!h]
\includegraphics[angle = 0, width = 0.9\columnwidth]{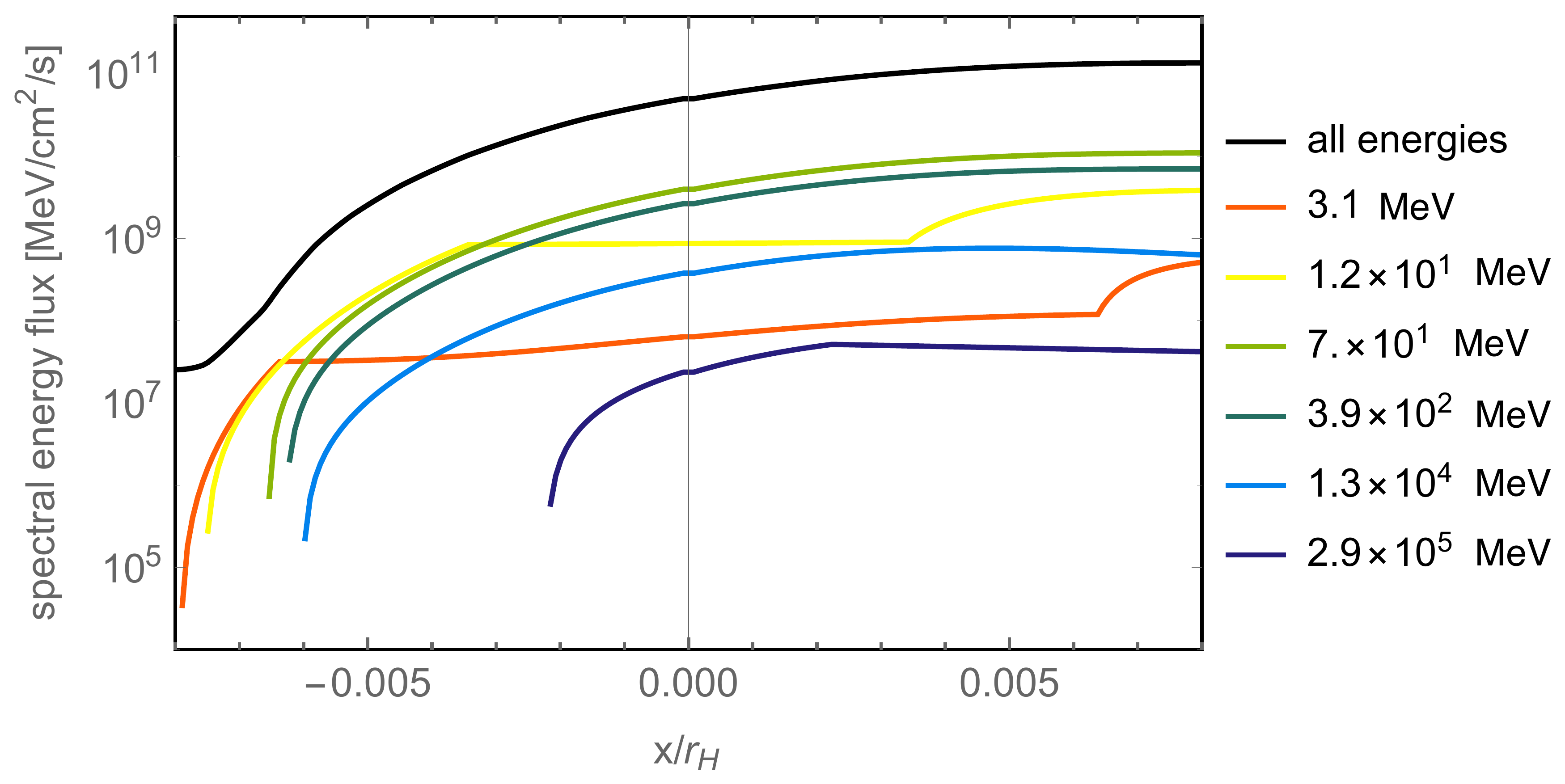}
\caption{The outgoing photon energy flux as a function of position inside of the gap. This result is for a maximumly spinning BH of mass $10^7 M_\odot$ with a magnetic field strength of $10^4$ Gauss and an ambient energy density of $10^6 \text{ ergs}/\text{cm}^3$. The $x$-axis has been normalized by the BH radius.}
\label{1d_flux}
\end{figure}

The solution for the structure of the gap for a $10^7 M_\odot$ maximumly spinning BH with an ambient photon energy density of $10^6 \text{ ergs}/\text{cm}^3$ and sitting in magnetic field of strength $10^4$ Gauss is shown in Figs. \ref{1d_GandE}-\ref{1d_flux}. Fig. \ref{1d_GandE} details the electric field in the gap and compares it to the Lorentz factor. The charges inside of the gap gain kinetic energy -- for which the Lorentz factor can be used as a proxy -- by being accelerated by the electric field and lose energy via inverse Compton scattering.
The charge densities produced by the cascade is shown in Fig. \ref{1d_pairs} and the outgoing photon energy flux is illustrated in Fig. \ref{1d_flux}. The outgoing photons are potentially observable, depending on the environment around the BH. Fig. \ref{1d_flux} shows the approximate peaked spectral energy at 70 MeV (the peak energy can be seen more clearly in Fig. \ref{Ubtrans}).

As shown in Eq. \ref{dE}, the 1D solution is for a particular $\rho_{GJ}$ at a given angle, $\theta$, with respect to the axis of rotation. By solving for $\rho_{GJ}$ as a function of $\theta$, we can obtain the 2D structure of the gap. 
\begin{figure}[!h]
\includegraphics[angle = 0, width = 0.9\columnwidth]{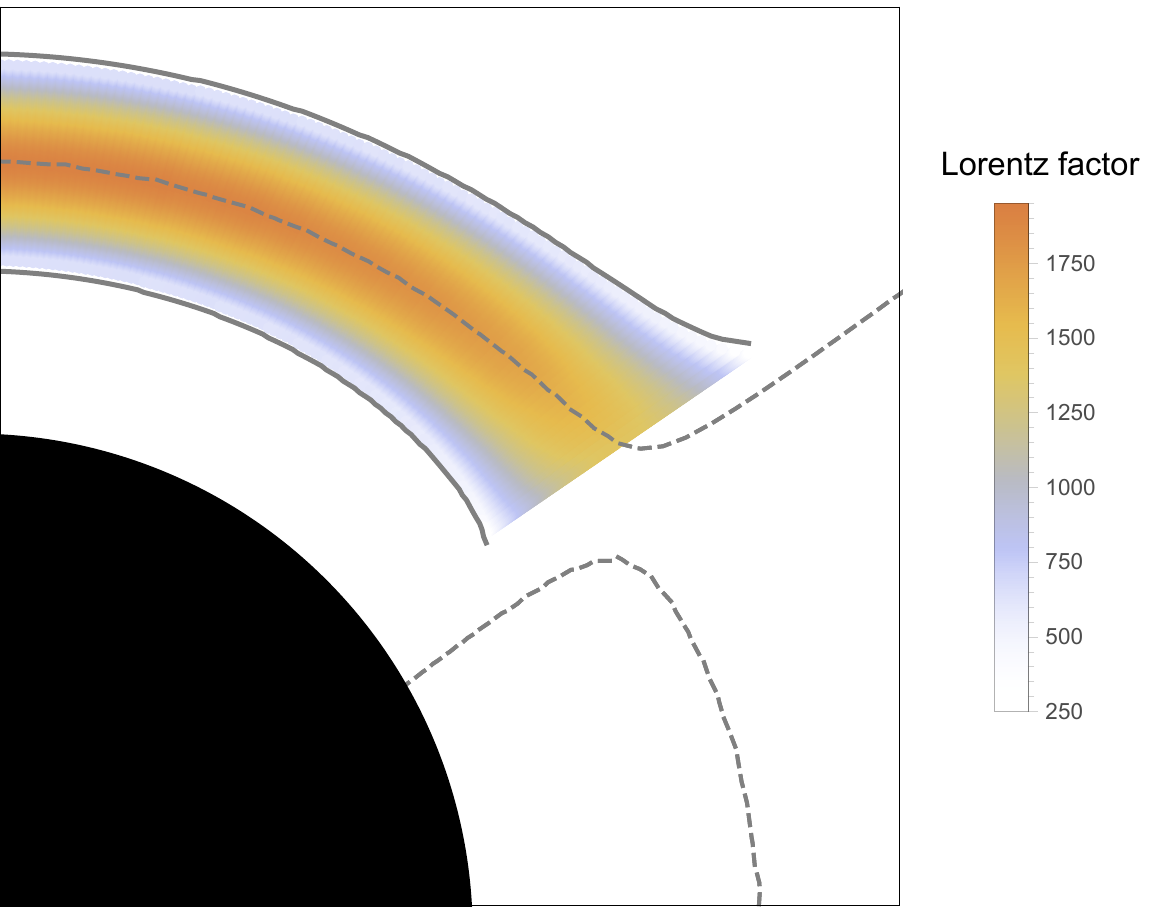}
\caption{Lorentz factor versus polar angle. This solution is for a BH mass of $10^7 M_\odot$ with a magnetic field strength of  $10^4$ Gauss and an ambient energy density of $10^6 \text{ ergs}/\text{cm}^3$}
\label{gamma2D}
\end{figure}
\begin{figure}[!h]
\includegraphics[angle = 0, width = 0.9\columnwidth]{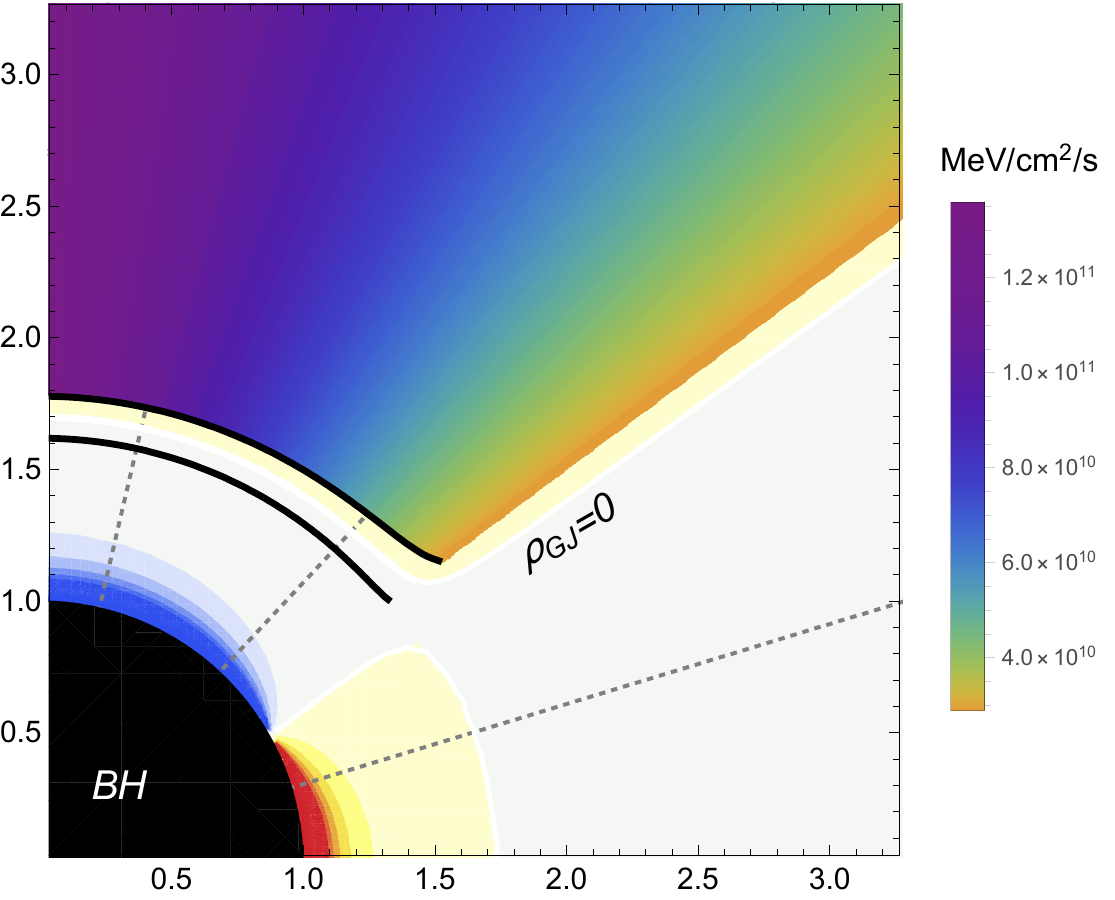}
\caption{The outgoing energy flux from the up-scattered photons as a function of polar angle. This solution is for a BH mass of $10^7 M_\odot$ with a magnetic field strength of  $10^4$ Gauss and an ambient energy density of $10^6 \text{ ergs}/\text{cm}^3$}.
\label{flux}
\end{figure}
To examine the efficiency of the cascade process we can use the gap width as a proxy. Similarly, we can use the Lorentz factor as a proxy for the available energy. The gap width in Fig. \ref{gamma2D} is not to scale but reflects the actual shape of the gap. By looking at Fig. \ref{gamma2D}, we can see that the cascade process is most efficient and energetic along the axis of rotation. Fig. \ref{flux} demostrates the outgoing energy flux of the $\gamma$-rays that are from the Comptonization of the ambient photons.

\section{Varying Physical Parameters}
The model has four parameters that can be varied: the mass and spin of the BH, the ambient magnetic field, and the background photon energy density. By changing the magnetic field, mass, or spin; the physical change to our model is in $\rho_{GJ}$. By changing the background energy density; the physical changes to our model are in the angle-averaged pair production redistribution function and the Compton redistribution function. By changing these parameters we can gain insight into how they effect the different aspects of the cascade process, \emph{i.e.}, the gap width, the peak Lorentz factor, maximum electric field, \emph{etc}.

\subsection{Changing the Goldreich-Julian Charge Density }

\subsubsection{Varying the Black Hole Mass}
 \begin{figure}[!h]
\includegraphics[angle = 0, width = 0.9\columnwidth]{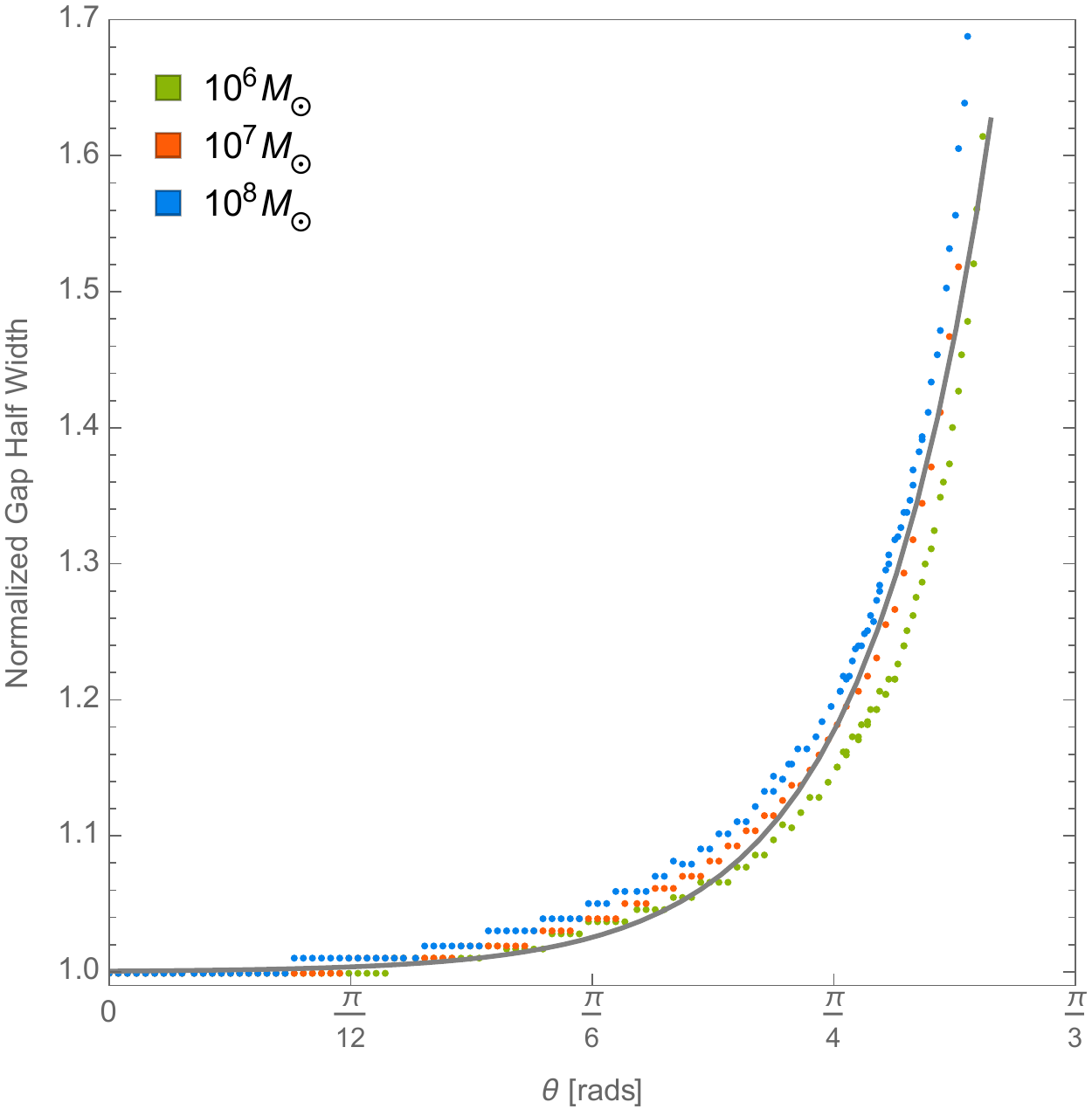}
\caption{The width of half of the gap normalized at the axis of rotation versus the polar angle. An exponential fit of all three masses is  $1+5.2\times10^{-4}e^{7.4\theta}$. This demonstrates that the efficiency of the cascade process as a function of polar angle, while the gap is thin, is invariant relative to the mass of the BH.}
\label{normmassgap}
\end{figure}
Changing the mass and observing how the structure of the gap changes, gives us insight into the conditions needed to produce AGN. Normalizing the gap width to one at the axis of rotation, Fig. \ref{normmassgap}, we can see that the relationship between the gap width and inclination angle is invariant with respect to the BH mass, while the gap is thin. The gap half width as a function of $\theta$ is $H\propto \text{const}+e^{7.4\theta}$. Hereafter, the $\theta$ dependence is valid for $0 \leq \theta < \theta_\text{max}=\cos ^{-1}\left\{1/\sqrt{3}\right\}$. Similar graphs can be made for the magnetic field, spin, and background energy density. The associated fits are provided in the Appendix and listed in Table \ref{tableall}.

\subsubsection{Varying the Magnetic Field}
The magnetic field that threads the BH likely originates from the environment, i.e. the accretion disk. 
\begin{figure}[!h]
\includegraphics[angle = 0, width = 0.9\columnwidth]{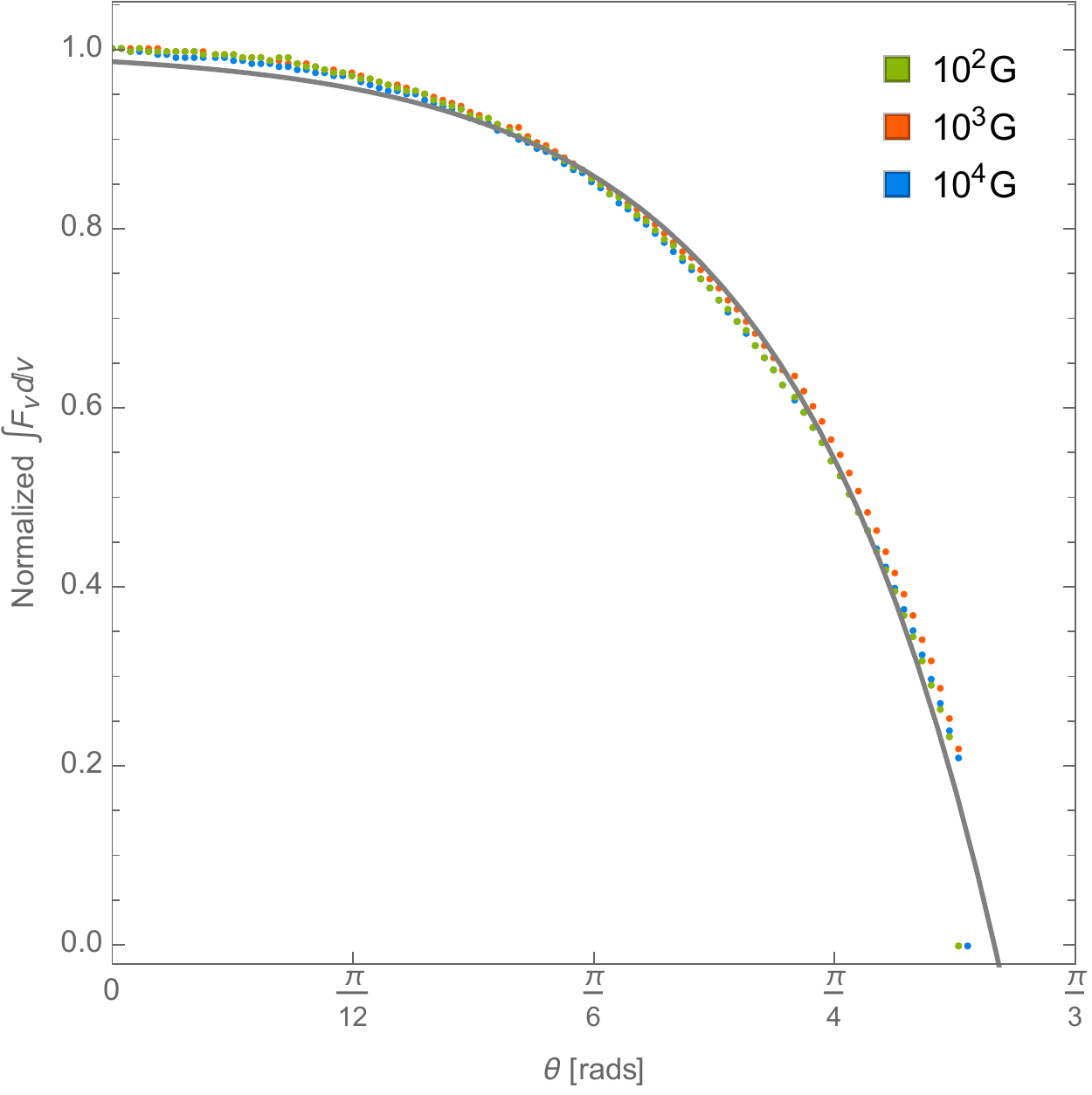}
\caption{The outgoing photon energy density normalized at the axis of rotation versus the polar angle. An exponential fit of all three magnetic fields is $1-0.013 e^{4.5 \theta}$.}
\label{Bnormspec}
\end{figure}
Fig. \ref{Bnormspec} shows the drop off in outgoing photon energy flux as a function of $\theta$ with the outgoing luminosity normalized to one at the axis of rotation, $\int F_\nu d\nu\propto \text{const}-e^{4.5\theta}$, for a sizable range of magnetic fields. As $\theta$ increase the outgoing energy flux drops by approximately 80\% at large $\theta$ ($\sim\theta_\text{max}$). Similar graphs can be made for the BH mass, spin, and background energy density. Those fits are provided in the Appendix and listed in Table \ref{tableall}.

\subsubsection{Spin}
\begin{figure}[!h]
\includegraphics[angle = 0, width = 0.9\columnwidth]{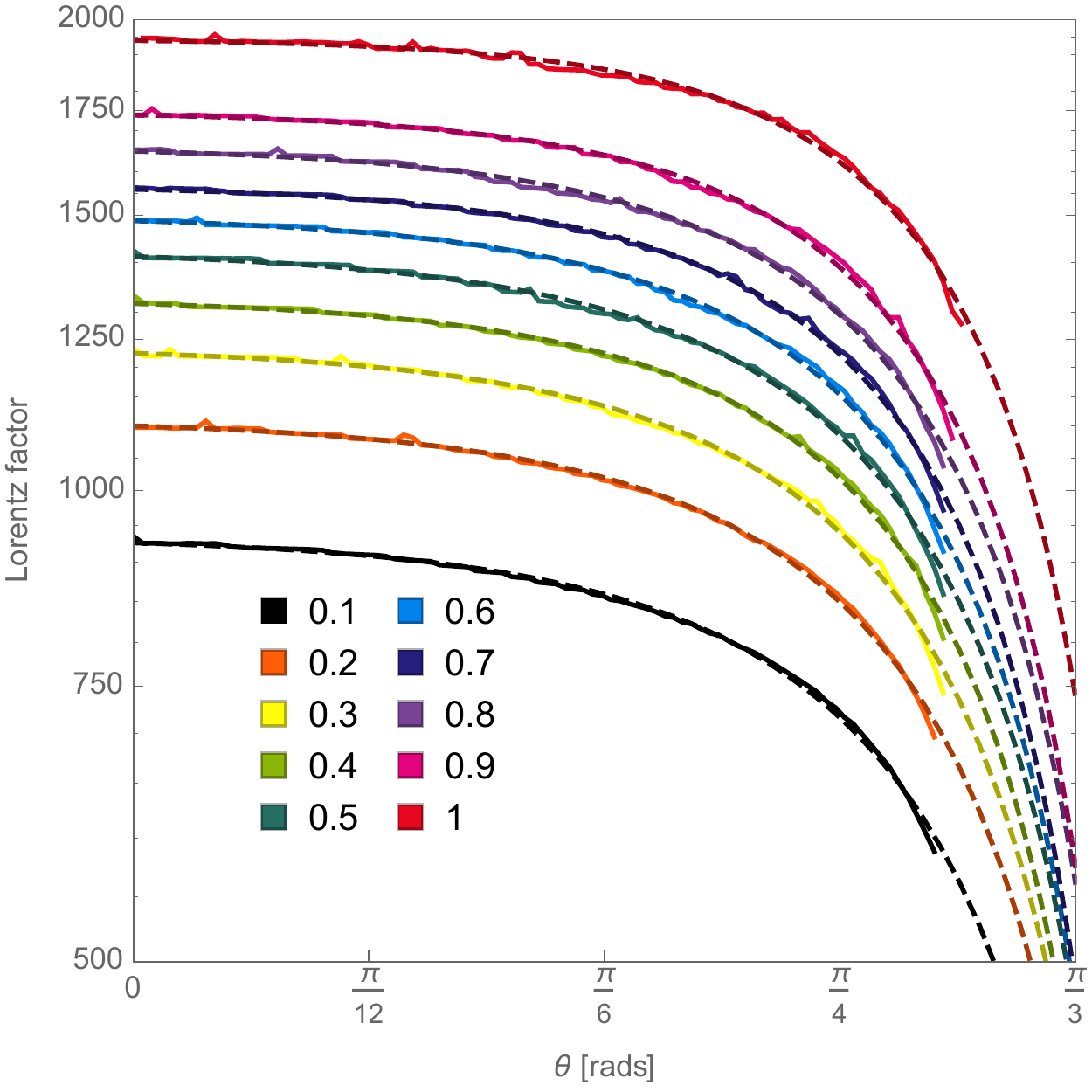}
\caption{Lorentz factor of the gap versus polar angle. The 10 curves represent the change in width as polar angle increase going away from the axis of rotation for 10 different spin and their corresponding fits represented with dashed lines. From the top down the spins are 1, 0.9, 0.8, 0.7, 0.6, 0.5, 0.4, 0.3, 0.2, and 0.1. And similarly, the fits from the top down are $1900-6.3 e^{5.0 \theta }$, $940 -9.5 e^{4.0 \theta }$, $1100 -11 e^{4.0 \theta }$, $1200 -13 e^{4.0 \theta }$, $1300 -12 e^{4.1 \theta }$, $1400 -15 e^{4.0 \theta }$, $1500 -14 e^{4.1 \theta }$, $1600 -11 e^{4.4 \theta }$, $1700 -14e^{4.2 \theta }$, and $1700 -10 e^{4.5 \theta }$. }
\label{spingamma}
\end{figure}

To illustrate how large the gap is with respect to the BH environment, Fig. \ref{spingap} has the gap for varying spin overlaid on simplified version of Fig. \ref{detail}. The gap width is increased by an order of magnitude for demonstration purposes. 
The energy stored in the kinetic energy of the charges can be seen in Fig. \ref{spingamma} and details how the maximum Lorentz factor varies as a function of theta for different spins. Overlaid on the results in Fig. \ref{spingamma} are exponential fits as a function of $\theta$ for $\theta < \theta_\text{max}$; the fits are summarized in Table \ref{tableall}. These fits are useful, for example, in estimating the change in available energy as a function of inclination angle. Similar graphs can be produced for the other parameters and the fits are listed in Table \ref{tableall}.

We can find a fit for the maximum Lorentz factor normalized to one at the axis of rotation as a function of $\theta$ for all spins, $\Gamma_\text{max}\propto \text{const}-e^{4.4 \theta}$; fits of the normalized maximum Lorentz factor as a function of inclination angle for the other parameters are listed in Table \ref{tableall}.

\begin{figure}[!h]
\includegraphics[angle = 0, width = 0.9\columnwidth]{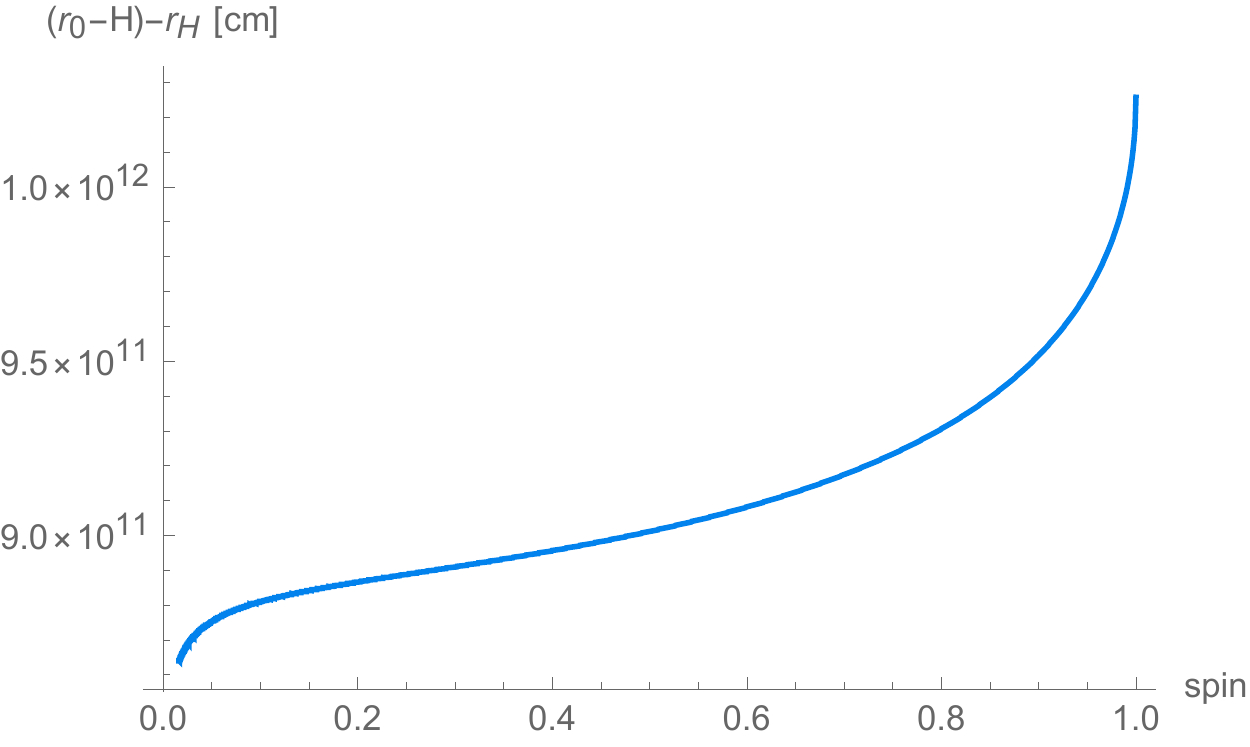}
\caption{The distance between the BH horizon and the inner edge of the gap as a function of spin for a BH of mass $10^7 M_\odot$ with a magnetic field strength of  $10^4$ Gauss and an ambient energy density of $10^6 \text{ ergs}/\text{cm}^3$.}
\label{inner2horizon}
\end{figure}

As spin decreases the BH's radius increases, this is shown in Eq. \ref{rH} and Fig. \ref{spingamma}. The interplay between the change in BH radius and gap width as a function of spin is shown in Fig. \ref{inner2horizon}. $r_0-H$ is the position of the inner edge of the gap and $(r_0-H)-r_H$ is the distance between the inner edge of the gap and the horizon of the BH. 
\begin{figure}[!h]
\includegraphics[angle = 0, width = 0.99\columnwidth]{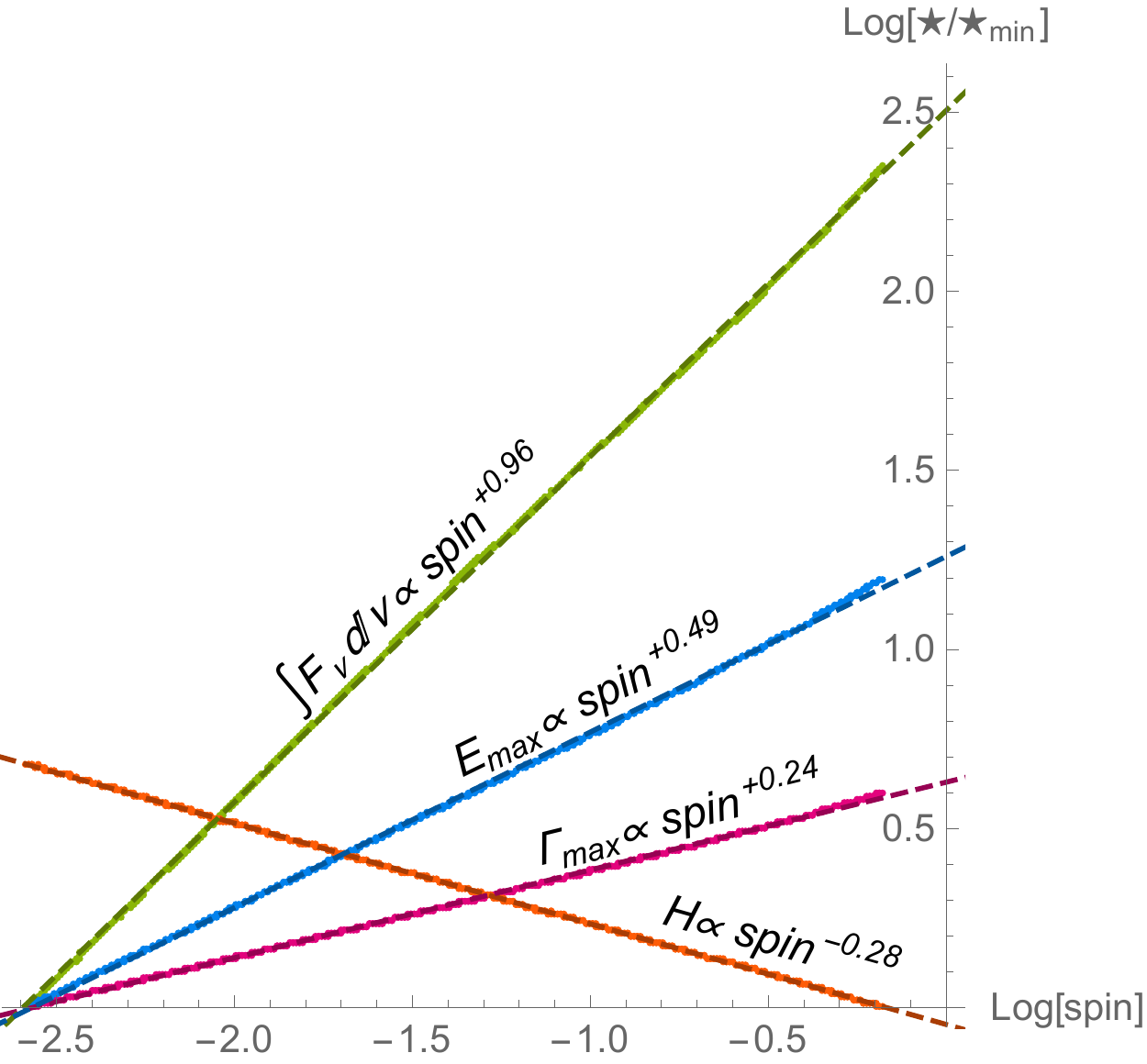}
\caption{The $\star$ is a place holder that represents the maximum Lorentz factor, maximum electric field, gap width, and photon energy flux. Each physical quantity is normalized to it's minimum value and then plotted with respect to the spin of the BH on a log-log scaling.}
\label{spinpropto}
\end{figure}

Probing the structure of the gap over all spin, allows us to obtain relationships between physical parameters and $a$. Fig. \ref{spinpropto} shows the gap half width, the maximum Lorentz factor, the maximum electric field, and the outgoing photon energy flux plotted with respect to $a$ on a log-log scaling after being normalized by their minimum value for the range in spin shown. These relationships allow estimations of the structure of the gap for any BH of mass $10^7$ embedded in a $10^4$ Gauss magnetic field with an available background photon energy density of $10^6$ ergs/cm$^3$ to be made. Similar graphs can be made for the magnetic field, BH mass, and background energy density. The associated fits are provided in the Appendix and the background energy density graph is shown in Fig. \ref{Ubpropto}.

\subsection{Changing Energy Available for Plasma Cascade }
\subsubsection{Background Energy Density}

\begin{figure}[!h]
\centering
\begin{tabular}{c}
\includegraphics[angle = 0, width = 0.9\columnwidth]{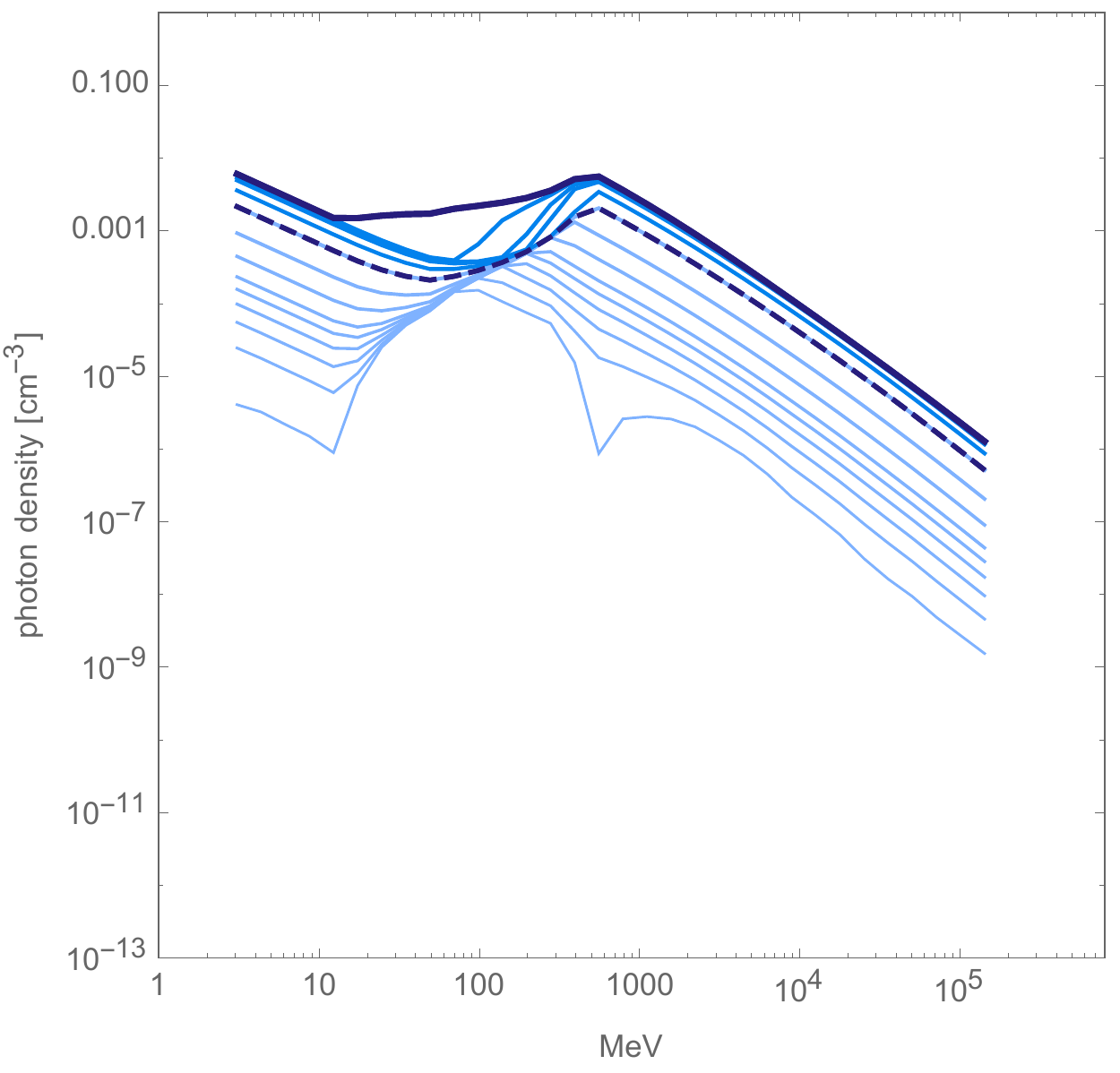}\\
\includegraphics[angle = 0, width = 0.9\columnwidth]{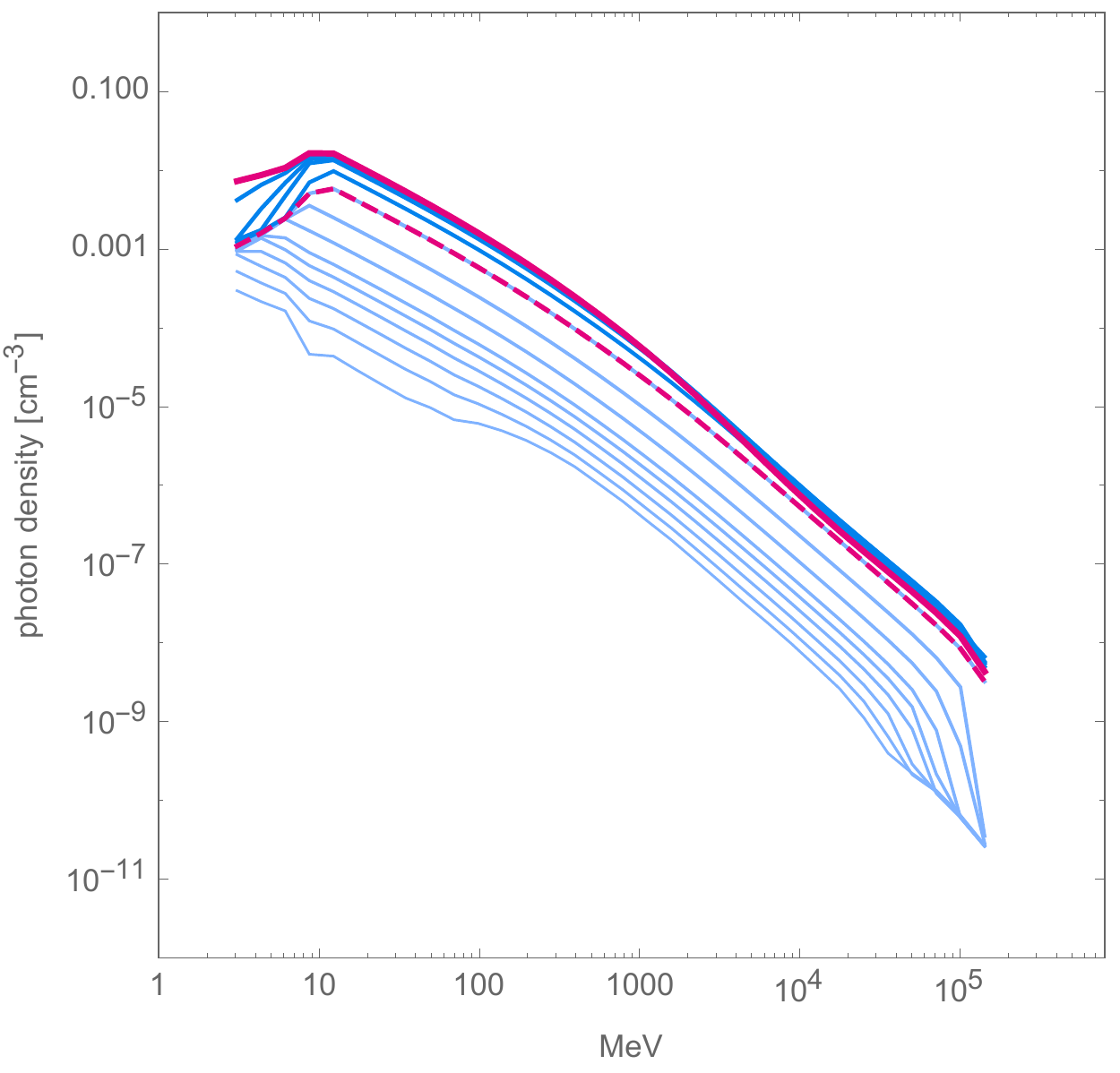}
\end{tabular}
\caption{A comparison of the spectral transition from the inner boundary of the gap through the center of the gap (bold dashed line) and to the outer boundary of the gap (bold solid line). Starting at 12\% of the gap width after the inner (closest to the BH) boundary (bottommost curve), 14 spectral lines are shown. The top spectral transition plot is for $U_b=10^5$ ergs/cm$^3$. The bottom spectral transition plot is for $U_b=10^6$ ergs/cm$^3$.  }
\label{Ubtrans}
\end{figure}

Changing the background photon energy density around the BH changes the energy available for $e^\pm$ to inverse Compton scatter with and for $\gamma$-rays to pair produce with in the model. Changing  $U_b$ and observing how the structure of the gap changes, gives us insight into the conditions needed to produce AGN. Fig. \ref{Ubtrans} provides a comparison of the spectral transition through the gap for two different background photon energy densities. Each line is a snapshot in space of the up-scattered spectrum in the gap. The top plot is for $U_b=10^5$ ergs/cm$^3$ and has a peak in it's spectrum around $10^3$ MeV. The bottom plot is for $U_b=10^6$ ergs/cm$^3$ is peaked around 10 MeV.
\begin{figure}[!h]
\includegraphics[angle = 0, width = 0.99\columnwidth]{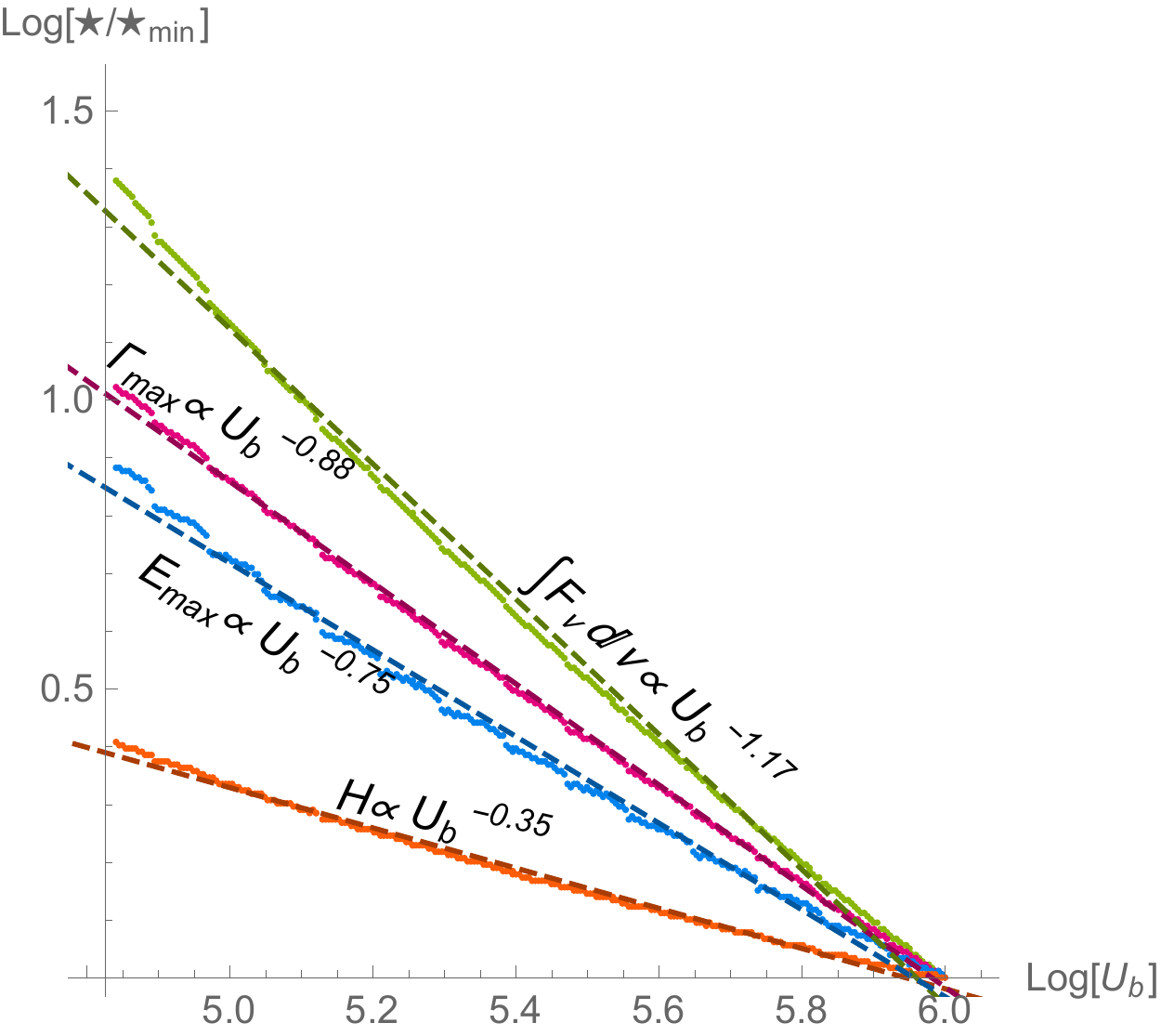}
\caption{The $\star$ is a place holder that represents the maximum Lorentz factor, maximum electric field, gap width, and photon energy flux. Each physical quantity is normalized to it's minimum value and then plotted with respect to the background energy density on a log-log scaling.}
\label{Ubpropto}
\end{figure}

\begin{figure*}
\begin{tabular}{cc}
	\includegraphics[width = 0.85\columnwidth]{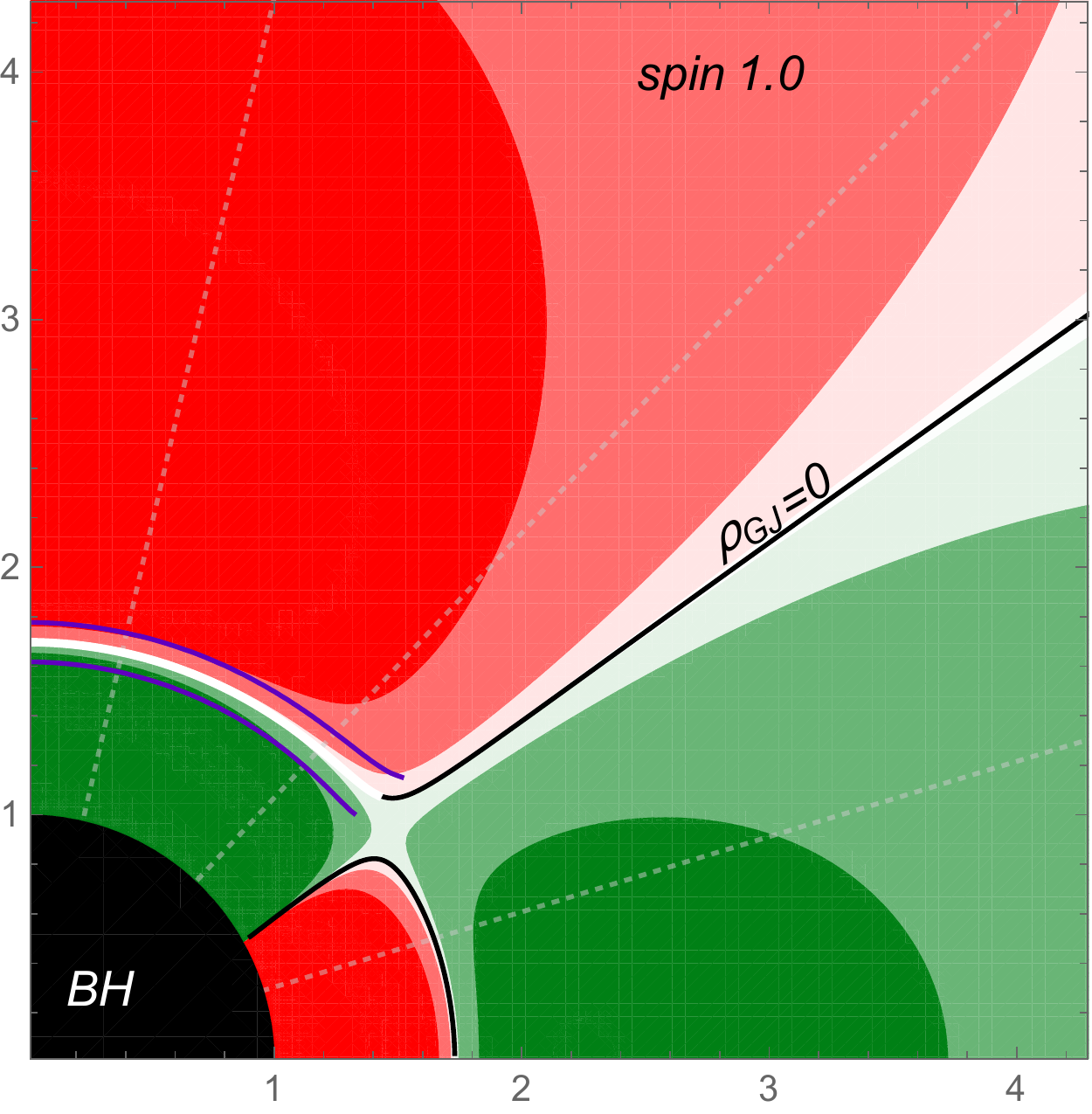} &  
	\includegraphics[width=0.85\columnwidth]{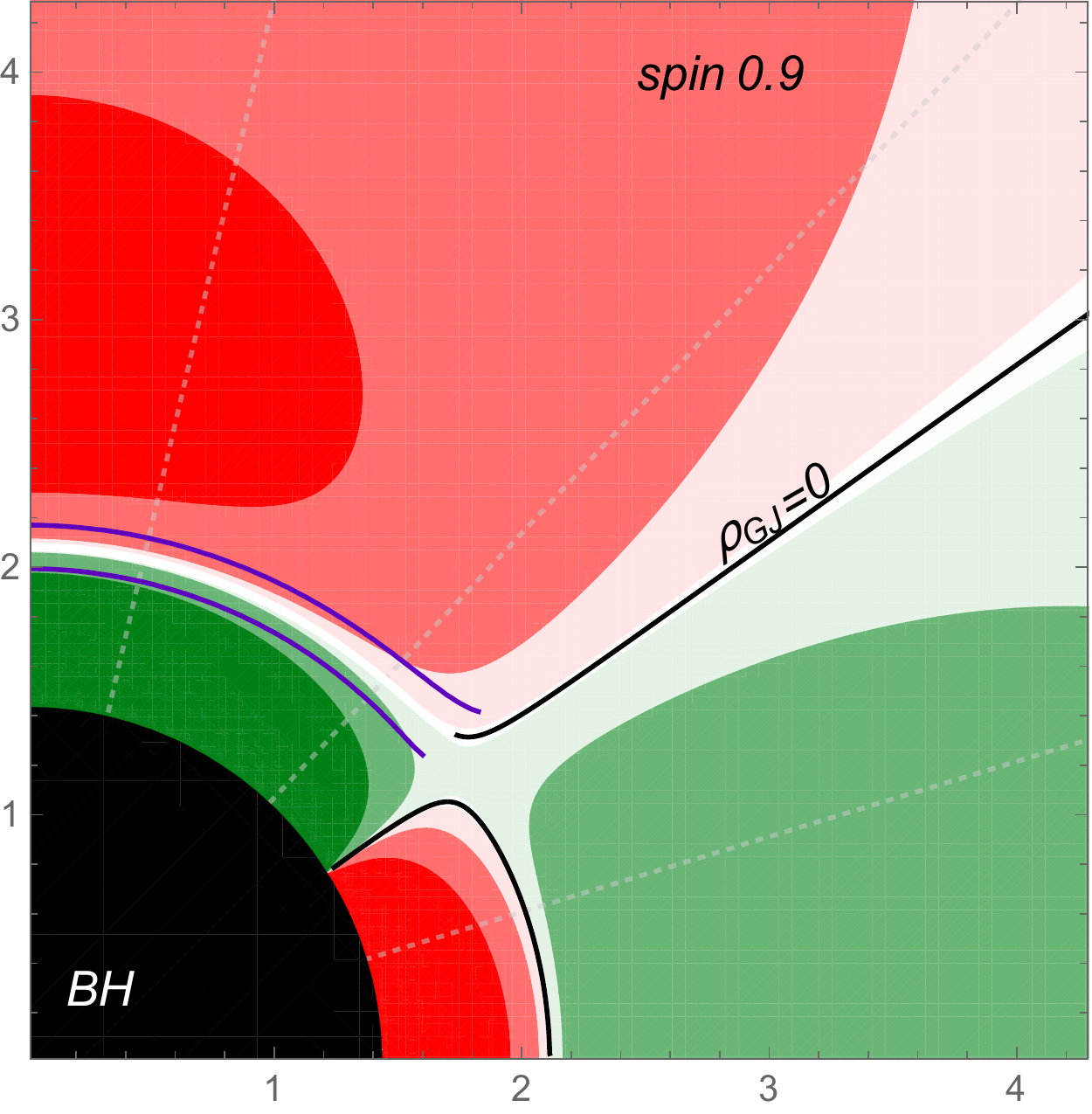} \\
	\includegraphics[width=0.85\columnwidth]{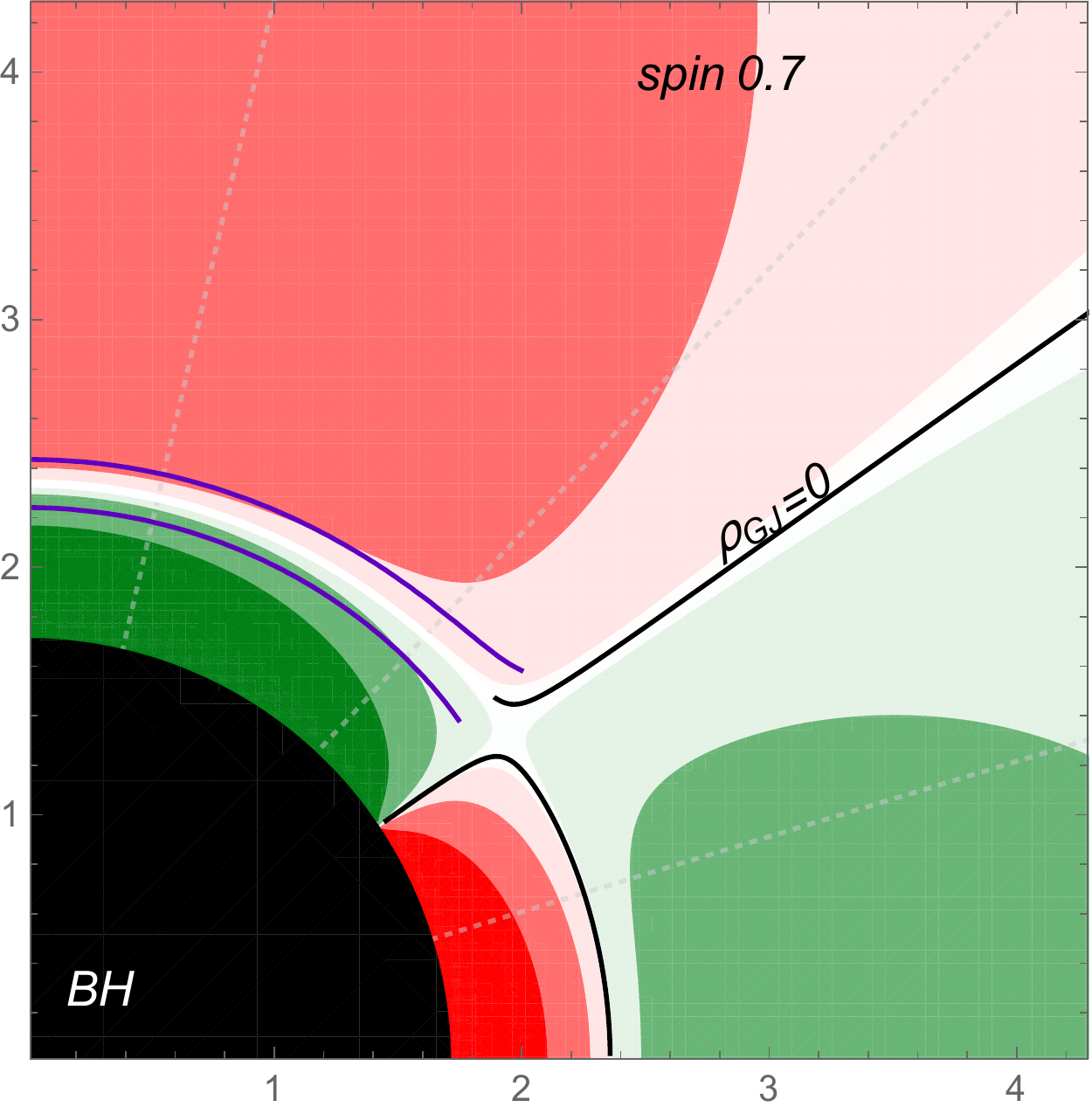} &  

	\includegraphics[width = 0.85\columnwidth]{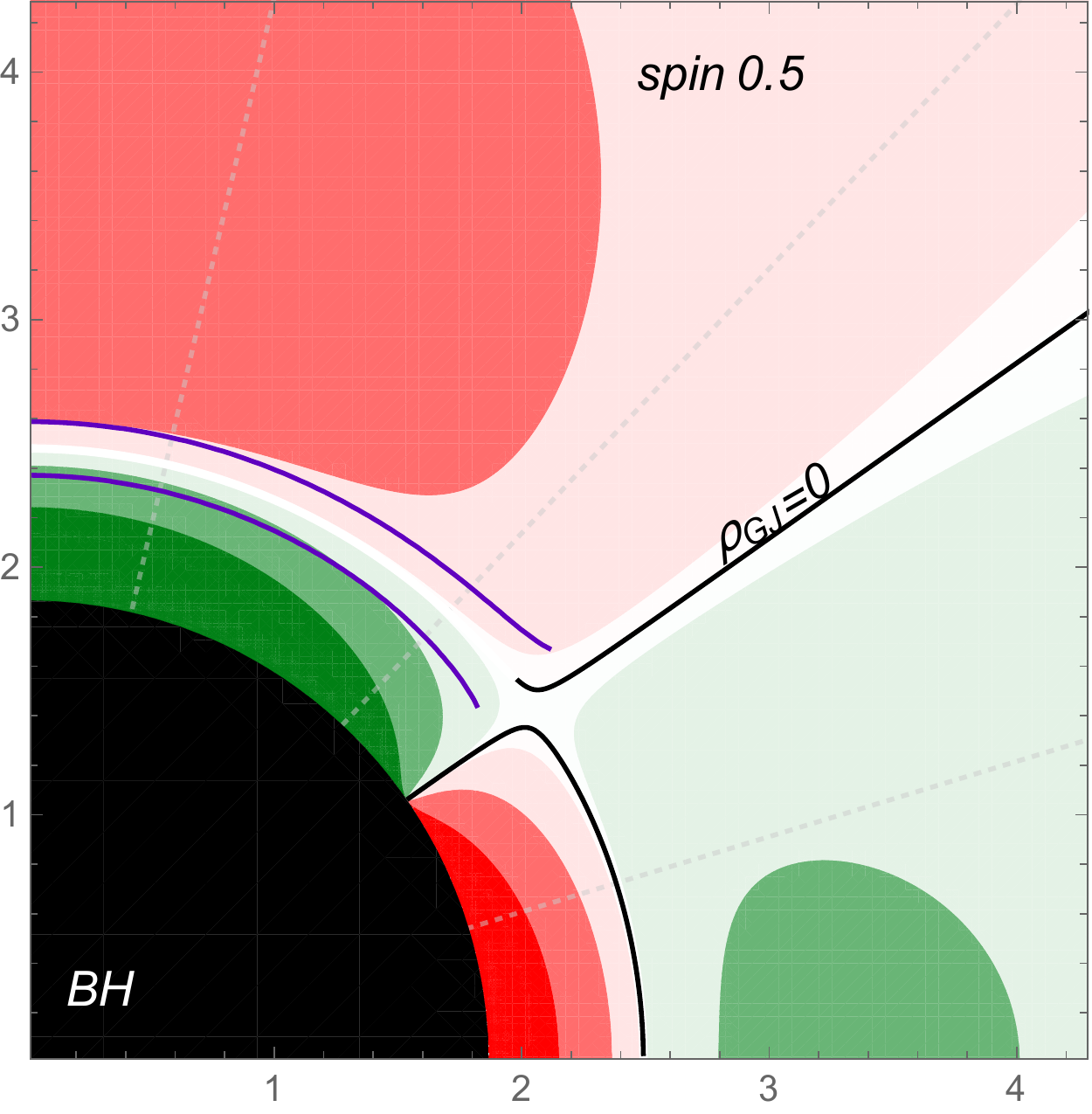} \\
	\includegraphics[width = 0.85\columnwidth]{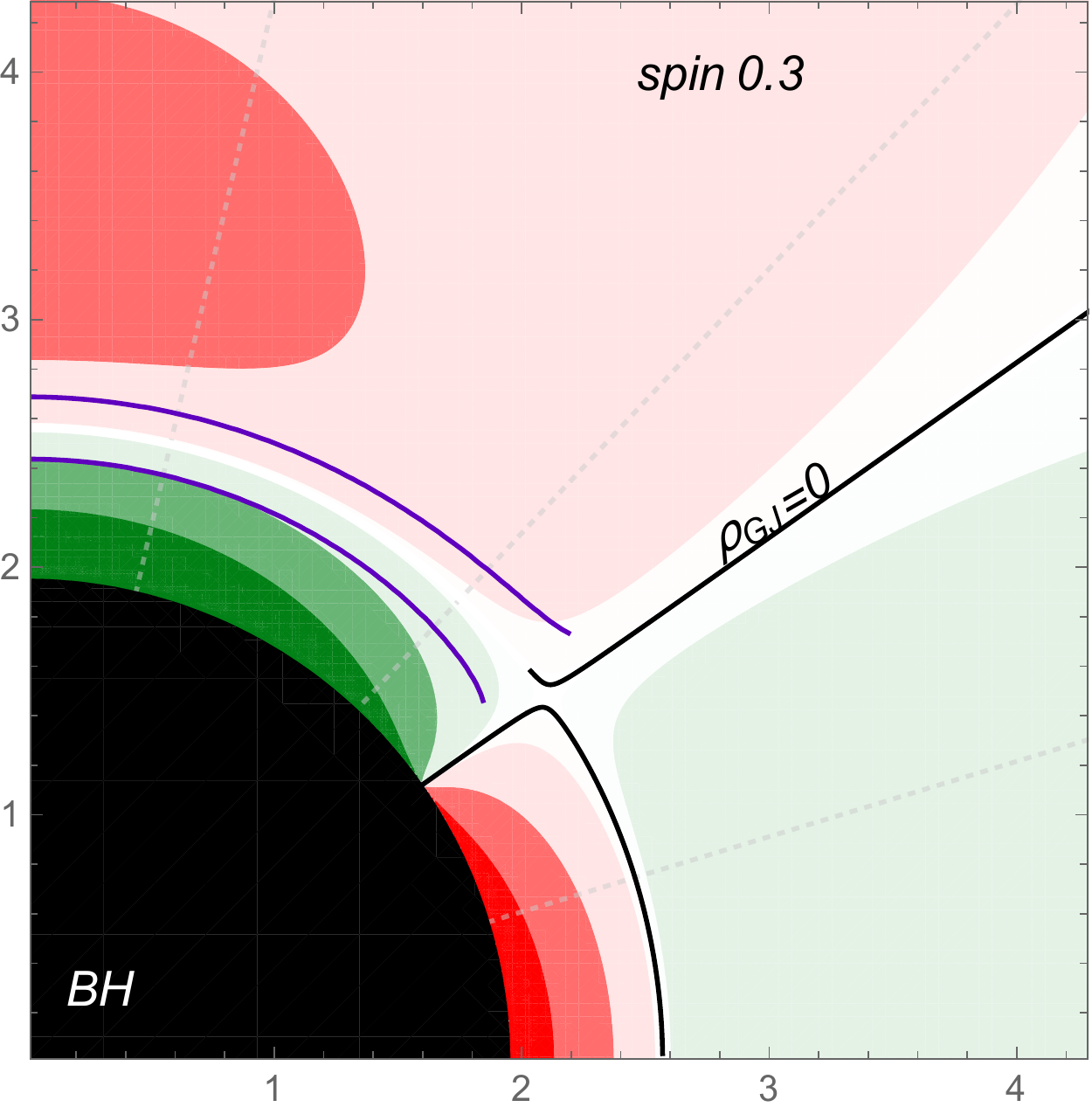} &
	\includegraphics[width = 0.85\columnwidth]{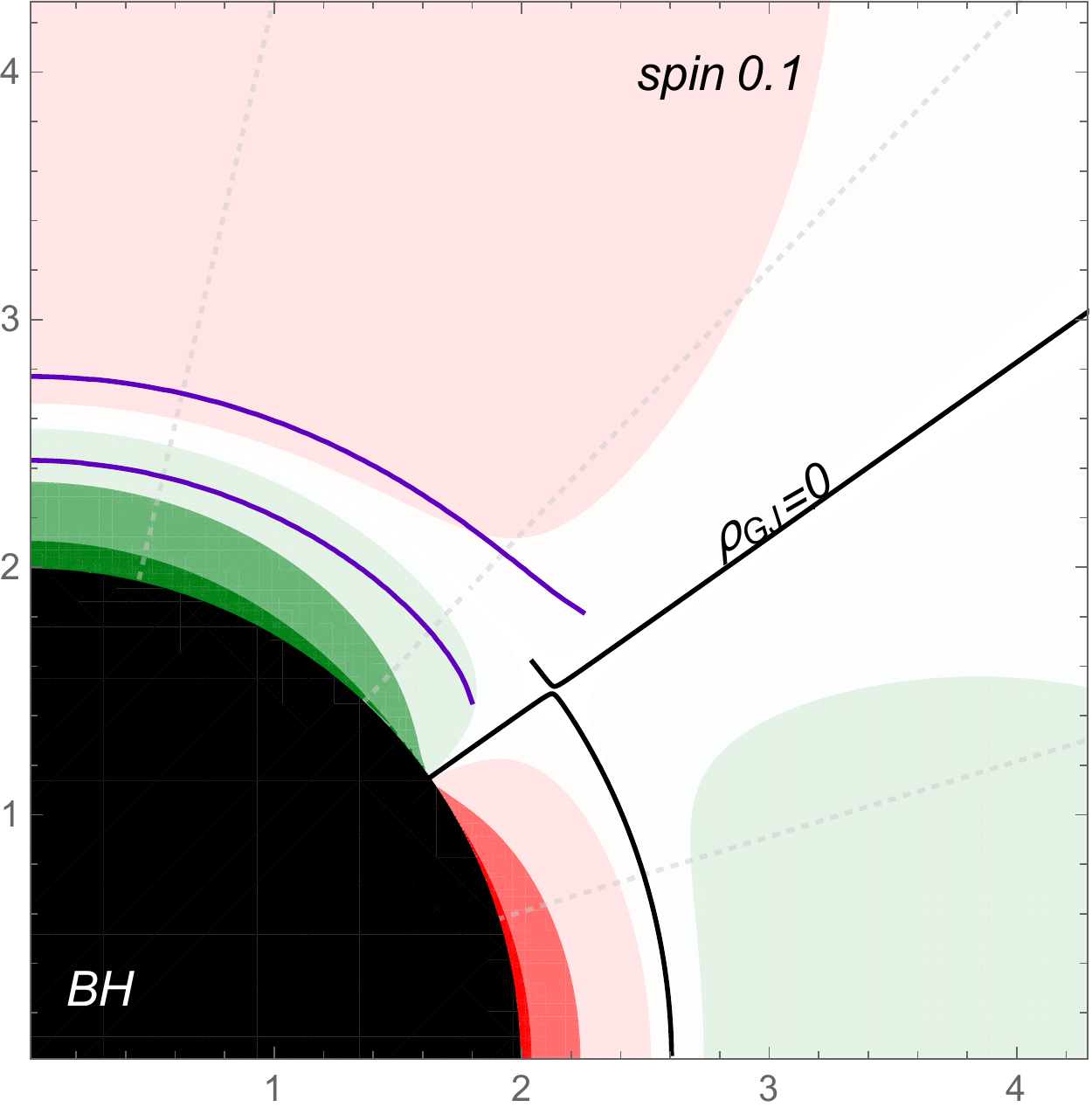} \\

\end{tabular}
\caption{Gap width increased by an order of magnitude. for a maximally spinning BH, the radius is set to one. The plasma density is displayed in red and green which correspond to positive and negative charge densities, respectively. It can be seen that as the BH's spin decreases the gap width increase and the plasma density around the gap decreases.}
\label{spingap}
\end{figure*}

After probing the structure of the gap over all spin, we can construct a plot similar to Fig. \ref{spinpropto}; Fig. \ref{Ubpropto} shows the gap half width, the maximum Lorentz factor, the maximum electric field, and the outgoing photon energy flux plotted with respect to the background energy density on a log-log scaling after being normalized by their minimum value over the shown range of $U_b$. These relationships allow us to estimate, for example, the gap width versus the BH radius for any maximumly spinning mass BH of mass $10^7 M_\odot$ embedded in a $10^4$ Gauss magnetic field.

\section{Discussion}
Combining the data shown in Figures \ref{normmassgap}, \ref{Bnormspec}, \ref{spinpropto}, and \ref{Ubpropto} and the data listed in Table \ref{tableall} we can construct expressions to estimate the structure of the gap for any mass, spin, magnetic field, and background energy density with an angular dependence. 

\begin{multline}
H(\theta)\simeq
1.1\times10^{10}\left\{1+2.8\times 10^{-3} e^{5.7 \theta }\right\} a^{-0.31}\times\\ \left[\frac{M}{10^7M_\odot}\right]^{0.54} \left[\frac{B}{10^4\text{Gauss}}\right]^{-0.27}\left[\frac{U_b}{10^6\text{ergs/cm$^3$}}\right]^{-0.22}\text{cm}.	
\label{hoftheta}
\end{multline}

\begin{multline}
\Gamma_\text{max}(\theta)\simeq
1.9\times10^{3}\left\{1-5.1\times 10^{-3} e^{4.6 \theta }\right\} a^{0.24} \times\\
\left[\frac{M}{10^7M_\odot}\right]^{-0.52} \left[\frac{B}{10^4\text{Gauss}}\right]^{0.25}\left[\frac{U_b}{10^6\text{ergs/cm$^3$}}\right]^{-0.88}	.
\label{goftheta}
\end{multline}

\begin{multline}
E_\text{max}(\theta)\simeq
69\left\{1-1.4\times 10^{-2} e^{4.1 \theta }\right\}a^{0.49}\times\\ 
\left[\frac{M}{10^7M_\odot}\right]^{-1.1} \left[\frac{B}{10^4\text{Gauss}}\right]^{0.49}\left[\frac{U_b}{10^6\text{ergs/cm$^3$}}\right]^{-0.75}\text{V/m}.
\label{eoftheta}	
\end{multline}

\begin{multline}
\int F_\nu d\nu(\theta)\simeq
6.7\times10^{14}\left\{1 -4.3\times10^{-2} e^{3.2 \theta}\right\}a^{0.96}\times\\
\left[\frac{M}{10^7M_\odot}\right]^{-4.5} \left[\frac{B}{10^4\text{Gauss}}\right]^{0.95}
\left[\frac{U_b}{10^6\text{ergs/cm$^3$}}\right]^{-1.17}\frac{\text{MeV}}{\text{cm$^2$s}}.
\label{fnuoftheta}	
\end{multline}

\begin{figure}[!h]
\includegraphics[angle = 0, width = 0.99\columnwidth]{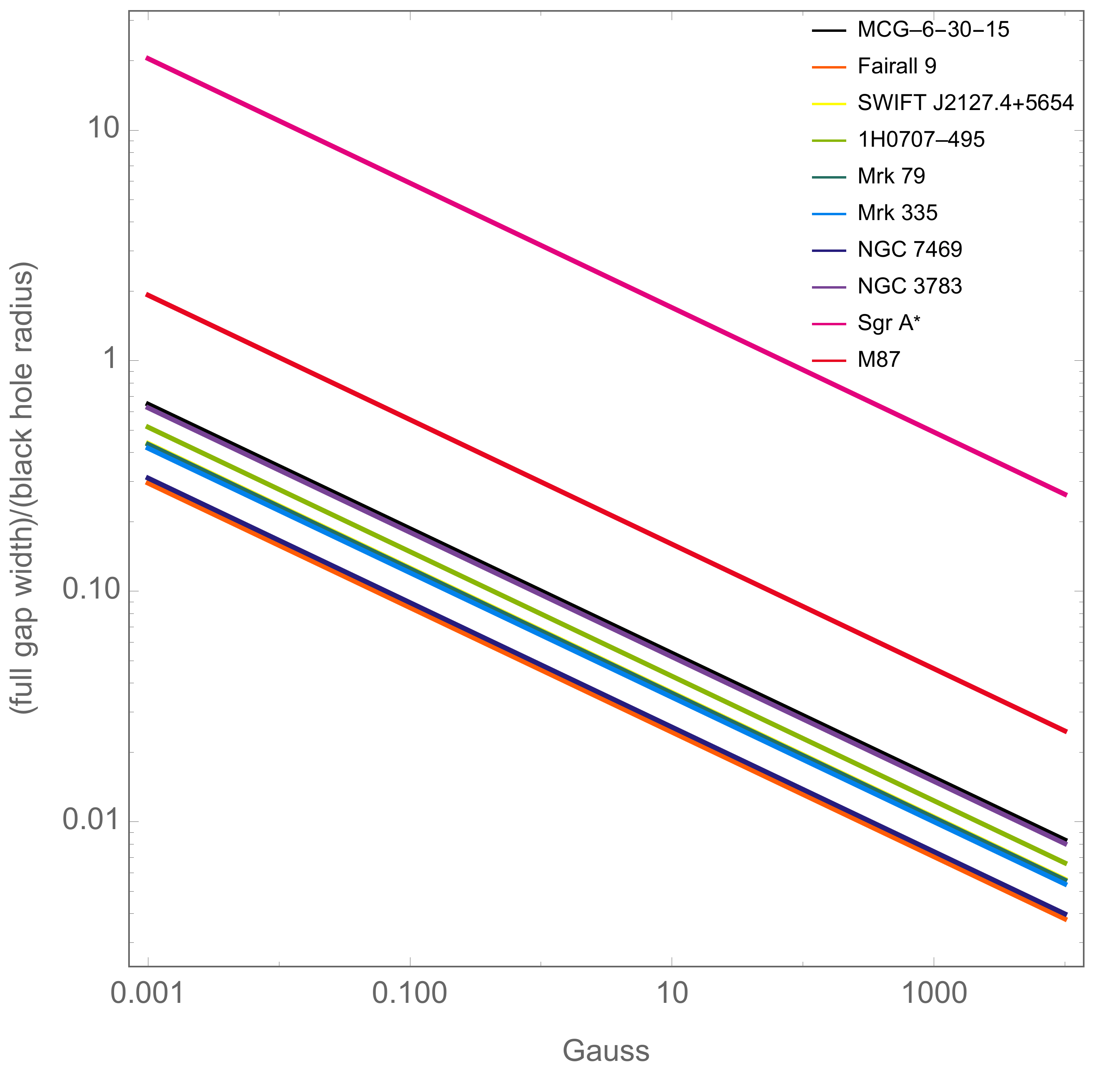}
\caption{The curve for Sgr $\text{A}^*$ is at the top and followed by M87. Next is MCG--6-30-15 and NGC 3783. They are followed by 1H0707-495. Next Mrk 79, Mrk 335, and SWIFT J2127.4+5654 are clustered together. They are followed by NGC 7469 and Fairall 9. The values for mass, spin, and energy density in Eq. \ref{hoftheta} are listed in Table \ref{numbersused}.}
\label{lotsofAGN}
\end{figure}

Using Eq. \ref{fnuoftheta}, the gap (inner jet) luminosity per steradian can be approximated,

\begin{multline}
\frac{dL}{d\Omega} \approx r_0(\theta)^2\times\int F_\nu d \nu (\theta)\simeq 2.5\times 10^{35}\times\\
\left\{1-4.3\times10^{-2} e^{3.2 \theta }\right\} \left\{1+1.6\times10^{-4} e^{5.0 \theta }\right\}^2 \times\\
 \left\{1+5.1\times10^{-3} e^{4.2 a}\right\}^2 a^{0.96}\left[\frac{M}{10^7M_\odot}\right]^{-2.5} \times
\\
\left[\frac{B}{10^4\text{Gauss}}\right]^{0.95}\left[\frac{U_b}{10^6\text{ergs/cm$^3$}}\right]^{-1.17}\text{ergs/s/sr},
\label{gaplum}	
\end{multline}
where we approximate the gap's radial distance in a similar fashion to Eqs. \ref{hoftheta}-\ref{fnuoftheta},
\begin{multline}
r_0(\theta)\simeq1.5\times 10^{13}\left\{1+1.6\times10^{-4}e^{5.0\theta}\right\}\times\\
\left\{1+5.1\times10^{-3}e^{4.2a}\right\}\left[\frac{M}{10^7M_\odot}\right]\text{cm}.
\end{multline}
By doubling Eq. \ref{hoftheta} and dividing by the radius of the BH, we can relate the relative full gap width, which is a proxy to the efficiency of the plasma cascade process over a wide range of parameters or for a particular object. Comparing an active galaxy, e.g., M87, to an inactive galaxy, e.g., Sagittarius A, is illustrative. First we must estimate the background energy density from the luminosity, $U_b\simeq L/(4\pi c r^2)$.
We can estimate $r$ using the innermost stable circular orbit \cite{Bardeen:1972fi},
\begin{equation}
r_\text{isco}=\frac{G M} {c^2}\left(3+Z_2-\sqrt{(3-Z_1) (Z_1+2 Z_2+3)}\right),
\label{isco}
\end{equation}
where
\begin{multline}
Z_1=\\
1+\left(\sqrt[3]{\frac{a c^2}{G M}+1}+\sqrt[3]{1-\frac{a c^2}{G M}}\right) \sqrt[3]{1-\frac{a^2 c^4}{G^2 M^2}},
\end{multline}
\begin{equation}
Z_2=\sqrt{\frac{3 a^2 c^2}{G M}+Z_1}.
\end{equation}
The luminosity of M87 is $2.7\times10^{42}$ ergs/s \cite{Prieto:2016eu}. Sgr $\text{A}^*$ has a luminosity of $10^{37}$ ergs/s \cite{Genzel:1994iy}. Let the ratio of the full width of the gap to the BH radius with $\theta=0$ be, 
\begin{multline}
\Upsilon=
9.4\times 10^8 r_H^{-1}a^{-0.31}\times\\
\left[\frac{M}{M_\odot}\right]^{0.54} \left[\frac{B}{\text{Gauss}}\right]^{-0.27}\left[\frac{L/(4\pi c r_\text{isco}^2)}{\text{ergs/cm$^3$}}\right]^{-0.22}	.	
\end{multline}	
Using a mass of $10^{9.5}M_\odot$, a spin of $0.65$, and a magnetic field of $15$ G for M87 yields $\Upsilon_\text{M87}=0.11$ \cite{1538-4357-676-2-L109, 0004-637X-786-1-5}. Similarly, using a mass of $10^{6.6} M_\odot$, a spin of $0.65$, and a magnetic field of $30$ G for Sgr $\text{A}^*$ yields $\Upsilon_\text{Sgr $\text{A}^*$}=1.3$ \cite{Johnson1242, Dokuchaev:2016epe}. The order of magnitude difference between $\Upsilon$ is consistent with M87 being active and Sgr $\text{A}^*$ not being active. Fig. \ref{lotsofAGN} displays the gap width over BH radius versus magnetic field strength for M87 and Sgr $\text{A}^*$ and eight additional AGN. Table \ref{numbersused} contains the physical quantities used. Fig. \ref{lotsofAGN} shows that the ratio of the gap width to BH radius for AGN is $<1$ for reasonable values of the magnetic field. 

\begin{table}[!h]
\centering
\renewcommand{\arraystretch}{2}
\begin{tabular}{|>{\centering\arraybackslash}m{3.2cm}||>{\centering\arraybackslash}m{0.7cm}|>{\centering\arraybackslash}m{1.3cm}|>{\centering\arraybackslash}m{2.7cm}|}
 \toprule
 \hline
    AGN & Spin & Mass & Energy Density \\
     \hline\hline
   
   M87 & $0.65$ & $10^{9.5}M_\odot$ & $0.33$ ergs/cm$^3$\\
     \hline
  Sgr $\text{A}^*$ & $0.65$& $10^{6.6}M_\odot$ & $2.1$ ergs/cm$^3$ \\
     \hline
    MCG–6-30-15 & $0.98$ & $10^{6.65}M_\odot$ & $3.8\times 10^7$ ergs/cm$^3$ \\
     \hline
    Fairall 9 & $0.65$ & $10^{8.41}M_\odot$ & $8.2\times10^4$ ergs/cm$^3$ \\
     \hline
     SWIFT J2127.4+5654 & 0.65 & $10^{7.18}M_\odot$ & $5.0\times10^6$ ergs/cm$^3$ \\
     \hline
     1H0707–495 & $0.98$ & $10^{6.7}M_\odot$ & $8.4\times 10^7$ ergs/cm$^3$ \\
     \hline
     Mrk 79 & $0.7$ & $10^{7.72}M_\odot$ & $4.0\times10^5$ ergs/cm$^3$ \\
     \hline
     Mrk 335 & $0.7$ & $10^{7.15}M_\odot$ & $7.5\times 10^6$ ergs/cm$^3$ \\
     \hline
     NGC 7469 & $0.69$ & $10^{7.09}M_\odot$ & $3.8\times 10^7$ ergs/cm$^3$ \\
     \hline
     NGC 3783 & $0.98$ & $10^{7.47}M_\odot$ & $8.5\times10^4$ ergs/cm$^3$ \\
     \hline
      \bottomrule
\end{tabular}
\renewcommand{\arraystretch}{1}	
\caption{The values used to in Eq. \ref{hoftheta} to make Fig. \ref{lotsofAGN} \cite{0004-637X-736-2-103, Prieto:2016eu, Genzel:1994iy, 1538-4357-676-2-L109, Dokuchaev:2016epe}.}
\label{numbersused}
\end{table}

\begin{figure}[!h]
\includegraphics[angle = 0, width = 0.99\columnwidth]{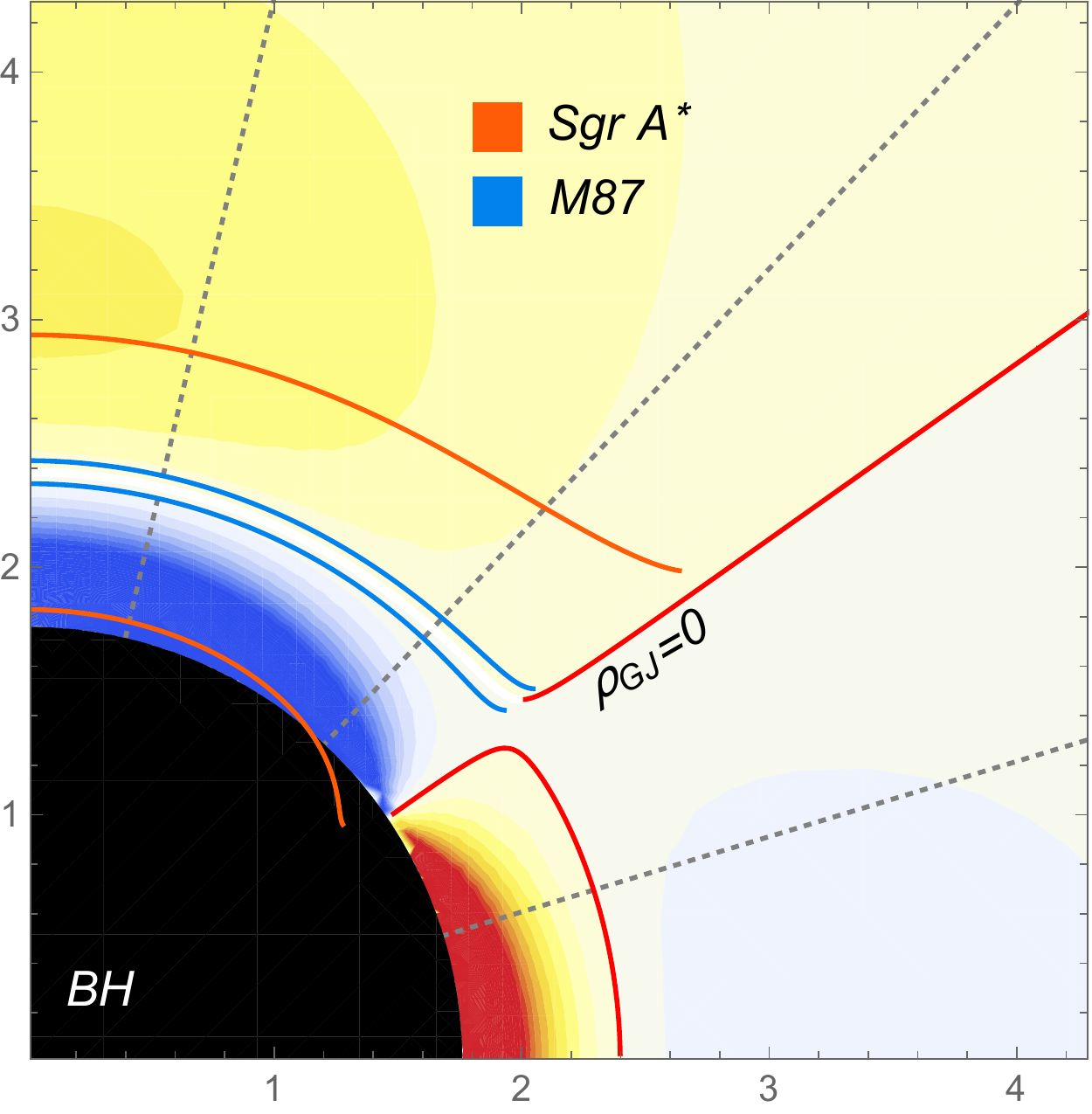}
\caption{The BH radius of a maximumly spinning BH has been set to one and the gap widths have been left to scale. M87 has a luminosity of  $2.7\times10^{42}$ ergs/s, mass of $10^{9.5}M_\odot$, a spin of $0.65$, and a magnetic field of $15$ G. Sgr $\text{A}^*$ has a luminosity of $10^{37}$ ergs/s, mass of $10^{6.6} M_\odot$, a spin of $0.65$, and a magnetic field of $30$ G.}
\label{M87vSgrA}
\end{figure}

Using the same example as above, M87 and Sgr $\text{A}^*$, we can use Eq. \ref{hoftheta} to make a plot similar to Fig. \ref{spingap}. Fig. \ref{M87vSgrA} shows the gap widths of M87 and Sgr $\text{A}^*$ to scale with the BH radius. The inner boundary of the gap of Sgr $\text{A}^*$ goes into the BH. Sgr $\text{A}^*$'s gap is too close to the event horizon to maintain the assumptions of symmetry in Eq. \ref{sym}. Further study is needed to get a clear understanding of the structure of the gap around Sgr $\text{A}^*$. However, a plausible interpretation of Fig. \ref{M87vSgrA} is that when the gap reaches the event horizon the cascade process becomes too inefficient and therefore the Blandford-Znajek process cannot power the jet. It is thus an intriguing possibility that this effect can explain the conditions needed for the AGN jet to occur.

\section{Conclusions}
 In this paper we explored a plasma cascade model that produces a force-free magnetosphere around stationary, axisymmetric Kerr BHs. A force-free magnetosphere is needed for the Blandford-Znajek mechanism to efficiently convert rotational energy from the BH into Poynting flux that can power relativistic jets. The 2D structure of the gap -- where the electron-positron cascade takes place -- in the magnetosphere was examined. Considering a gap that is thin with respect to the size of the BH, we assumed the structure inside of the gap to be symmetric and employed a power law spectrum with a single power law index. Using these assumptions we were able to numerically solve for the 2D structure of the gap over three orders of magnitude in BH mass and magnetic field, two orders of magnitude in background photon energy density, and over all spin. Probing this parameter space allowed us to construct Eqs. \ref{hoftheta}-\ref{fnuoftheta} and to give estimates for the structure of the gap for observed BHs (see Figs. \ref{lotsofAGN} \& \ref{M87vSgrA}). Solving for the 2D structure of the gap shows that the cascade is most efficient and energetic along the axis of rotation. A key aspect of the cascade process is the Comptonization of background photons; the outgoing energy flux of these photons and the gap (inner jet) luminosity can be estimated with Eqs. \ref{fnuoftheta} \& \ref{gaplum} and examples are shown in Figs. \ref{1d_flux} \& \ref{flux}. One intriguing observation, shown in Fig. \ref{M87vSgrA}, is that for non-jet-producing AGN, the distance between the inner edge of the gap and the BH horizon is small or vanishes. Bright AGN radio emission is commonly attributed to a jet. Here we showed that the pair- production efficiency controls the gap size and, likely, the origin of a jet. Thus we speculate that it is the gap cascade efficiency that can control and lead to the observed radio-loud/radio-quiet dichotomy of AGNs \cite{Sikora07}. Further investigation into the relationship between gap width and jet production is needed. All specifics of these results are tentative until further work has been done to use a more realistic background spectrum and relax the assumptions of symmetry in the gap. Nevertheless, the assumptions used do cover a large set of BH environments; and therefore, the general trends are expected to hold.

 \acknowledgments
MM is grateful to the the Institute for Theory and Computation at Harvard University for support and hospitality. The authors acknowledge DOE partial support via grant DE-SC0016368.

\clearpage

\appendix*

\section{Exponential Parameter Fits}
By probing the structure of the gap over several orders of magnitude in BH mass, we can find relationships between relevant physical parameters and the BH mass. These relationships allow us to estimate the energy output, energy available, cascade efficiency, etc. for any maximumly spinning BH embedded in a $10^4$ Gauss magnetic field with an available background photon energy density of $10^6$ ergs/cm$^3$
\begin{equation}
\begin{split}
	H=3.2\times10^6\left[\frac{M}{M_\odot}\right]^{0.51} \text{cm}, \\
	\\
	\Gamma_\text{max}=9.2\times 10^6 \left[\frac{M}{M_\odot}\right]^{-0.52},\\
	\\
	E_\text{max}=3.5\times10^9\left[\frac{M}{M_\odot}\right]^{-1.05} \text{V/m},\\
	\\
	\int F_\nu d\nu=2.0\times 10^{46}\left[\frac{M}{M_\odot}\right]^{-4.5}\text{MeV/cm$^2$/s}.\\
\end{split}
\end{equation}

Likewise, probing the structure of the gap for a range in background energy density , we can find relationships between relevant physical parameters and $U_b$. These relationships allow us to estimate the structure of the gap for any maximumly spinning BH of mass $10^7 M_\odot$ embedded in a $10^4$ Gauss magnetic field. The fits in Eq. \ref{Ubfits} are shown in Fig. \ref{Ubpropto}
\begin{equation}
\begin{split}
	H=1.4\times 10^{12}\left[\frac{U_b}{\text{ergs/cm$^3$}}\right]^{-0.35} \text{cm}, \\
	\\
	\Gamma_\text{max}=3.4\times 10^8\left[\frac{U_b}{\text{ergs/cm$^3$}}\right]^{-0.88},\\
	\\
	E_\text{max}=4.7\times 10^6\left[\frac{U_b}{\text{ergs/cm$^3$}}\right]^{-0.75} \text{V/m},\\
	\\
	\int F_\nu d\nu=1.8\times 10^{29}\left[\frac{U_b}{\text{ergs/cm$^3$}}\right]^{-1.2}\text{MeV/cm$^2$/s}.\\
	\end{split}
\label{Ubfits}
\end{equation}

Similarly we can find relationships between relevant physical parameters and the ambient magnetic field. These relationships allow us to estimate the structure of the gap for any maximumly spinning BH of mass $10^7 M_\odot$ with an available background photon energy density of $10^6$ ergs/cm$^3$
\begin{equation}
\begin{split}
	H=1.5\times 10^{11}\left[\frac{B}{\text{Gauss}}\right]^{-0.27} \text{cm}, \\
	\\
	\Gamma_\text{max}=190\left[\frac{B}{\text{Gauss}}\right]^{0.25},\\
	\\
	E_\text{max}=1.6\left[\frac{B}{\text{Gauss}}\right]^{0.49} \text{V/m},\\
	\\
	\int F_\nu d\nu=2.1\times 10^7\left[\frac{B}{\text{Gauss}}\right]^{0.95}\text{MeV/cm$^2$/s}.\\
\end{split}
\end{equation}

Finally, analyzing the structure of the gap over all spin, we can obtain relationships between physical parameters and $a$. These relationships allow us to estimate the structure of the gap for any BH of mass $10^7$ embedded in a $10^4$ Gauss magnetic field with an available background photon energy density of $10^6$ ergs/cm$^3$. The fits in Eq. \ref{spinfits} are shown in Fig. \ref{spinpropto}
\begin{equation}
\begin{split}
	H=1.3\times 10^{10}a^{-0.28} \text{cm}, \\
	\Gamma_\text{max}=1.6\times 10^3 a^{0.24},\\
	E_\text{max}=110 a^{0.49} \text{V/m},\\
	\int F_\nu d\nu=8.3\times 10^{10} a^{0.96}\text{MeV/cm$^2$/s}.\\
\end{split}
\label{spinfits}
\end{equation}

Table \ref{tableall} displays fits for the maximum Lorentz factor, half width of the gap and outgoing photon energy flux as a function of inclination angle for various BH masses and spins, ambient magnetic fields, and background spectral energy densities.

%%%%%%%%%%%%%%%%

\begin{table*}
\centering
\renewcommand{\arraystretch}{2}
\begin{tabular}{|>{\centering\arraybackslash}m{1.4cm}|>{\centering\arraybackslash}m{3.6cm}|>{\centering\arraybackslash}m{4.8cm}|>{\centering\arraybackslash}m{4.9cm}|}
 \toprule
 \hline
    Mass & Lorentz factor & Gap Half Width [cm] & Energy Flux [MeV/cm$^2$/s]\\
     \hline\hline
   
   $10^6 M_\odot$ & $7.0\times 10^3-11e^{5.9\theta}$ & $4.5\times 10^5e^{9.0\theta}+4.1\times10^{9}$ & $1.5\times 10^{13}-4.5\times 10^{11} e^{3.6 \theta}$\\
     \hline
  $10^7 M_\odot$  & $1.9\times 10^3-6.3e^{5.0\theta}$ & $7.1\times 10^6e^{7.2\theta}+1.2\times10^{10}$ & $1.4\times 10^{11}-2.8\times 10^9 e^{4.0 \theta}$\\
     \hline
     $10^8 M_\odot$  & $6.1\times 10^2-1.6e^{5.2\theta}$ & $1.3\times 10^7e^{8.1\theta}+4.1\times 10^{10}$ & $1.8\times 10^9-4.4\times 10^7 e^{3.8 \theta}$\\
     \hline
    all & $1-2.5\times10^{-3}e^{5.4\theta}$ & $5.2\times10^{-4}e^{7.4\theta}+1$ &  $1-1.7\times10^{-2}e^{4.2\theta}$\\
     \hline
      \bottomrule
   \toprule
\hline
    Magnetic Field & Lorentz factor & Gap Half Width [cm] & Energy Flux [MeV/cm$^2$/s]\\
     \hline\hline
   
   $10^2$ G & $2.0\times 10^{3}-6.3e^{5.0\theta}$ &$4.2\times 10^7e^{6.8\theta}+4.1\times 10^{10}$ & $1.7\times 10^9-3.3\times 10^7 e^{4.1 \theta}$\\
     \hline
  $10^3$ G &  $1.0\times10^{3}-2.2e^{5.4\theta}$ & $1.4\times 10^7e^{7.3\theta}+2.2\times10^{10}$ & $1.6\times 10^{10}-2.7\times 10^8 e^{4.2 \theta}$\\
     \hline
     $10^4$ G & $6.1\times10^{2}-1.4e^{5.3\theta}$ & $7.1\times 10^6e^{7.2\theta}+1.2\times10^{10}$ & $1.4\times 10^{11}-2.8\times 10^9 e^{4.0 \theta}$\\
     \hline
    all & $1-2.3\times10^{-3}e^{6.8\theta}$ & $1.0\times10^{-4}e^{5.3\theta}+1$ &  $1-0.013 e^{4.5 \theta}$\\
     \hline
      \bottomrule
\toprule
 \hline
    Energy Density & Lorentz factor & Gap Half Width [cm]& Energy Flux [MeV/cm$^2$/s]\\
     \hline\hline
   
   $10^5\frac{\text{ergs}}{\text{cm}^3}$ & $1.4\times 10^{4}-40e^{5.4\theta}$ &$1.2\times 10^7e^{7.4\theta}+2.6\times 10^{10}$ & $1.9\times 10^{12}-6.3\times 10^{10} e^{3.6 \theta}$\\
     \hline
  $10^6\frac{\text{ergs}}{\text{cm}^3}$ & $1.9\times10^{3}-6.3e^{5.0\theta}$ & $7.1\times 10^6e^{7.2\theta}+1.2\times10^{10}$ & $1.0\times 10^{10}-8.4\times 10^8 e^{2.6 \theta}$\\
     \hline
    all & $1-2.9\times10^{-3}e^{5.2\theta}$ & $7.6\times10^{-4}e^{6.8\theta}+1$ & $1-0.017 e^{4.3 \theta}$\\
     \hline
      \bottomrule
\toprule
 \hline
    Spin & Lorentz factor & Gap Half Width [cm] & Energy Flux [MeV/cm$^2$/s]\\
     \hline\hline
   
   1 & $1900-6.3 e^{5.0 \theta }$ &$7.1\times 10^6 e^{7.3 \theta }+1.2\times 10^{10}$ & $1.4\times 10^{11}-2.8\times 10^9 e^{4.0 \theta}$\\
     \hline
  0.9 & $940 -9.5 e^{4.0 \theta }$ & $1.7\times 10^7 e^{6.6 \theta }+1.3\times 10^{10}$ & $1.0\times 10^{10}-8.4\times 10^8 e^{2.6 \theta}$\\
     \hline
    0.8 & $1100 -11 e^{4.0 \theta }$ & $2.7\times 10^7 e^{6.3 \theta }+1.4\times 10^{10}$ & $1.9\times 10^{10}-1.6\times 10^9 e^{2.6 \theta}$\\
     \hline
    0.7& $1200 -13 e^{4.0 \theta }$ & $3.4\times 10^7 e^{6.1 \theta }+1.5\times 10^{10}$ & $5.6\times 10^{10}-4.0\times 10^9 e^{2.8 \theta}$\\
     \hline
      0.6& $1300 -12 e^{4.1 \theta }$ & $3.7\times 10^7 e^{6.1 \theta }+1.5\times 10^{10}$ & $4.6\times 10^{10}-3.4\times 10^9 e^{2.8 \theta}$\\
     \hline
      0.5& $1400 -15 e^{4.0 \theta }$ & $4.3\times 10^7 e^{6.1 \theta }+1.6\times 10^{10}$ & $3.7\times 10^{10}-3.1\times 10^9 e^{2.6 \theta}$\\
     \hline
      0.4& $1500 -13 e^{4.1 \theta }$ & $4.7\times 10^7 e^{6.1 \theta }+1.7\times 10^{10}$ & $2.8\times 10^{10}-2.2\times 10^9 e^{2.7 \theta}$\\
     \hline
      0.3 & $1600 -11 e^{4.4 \theta }$ & $5.1\times 10^7 e^{6.1 \theta }+1.9\times 10^{10}$ & $6.8\times 10^{10}-4.5\times 10^9 e^{2.9 \theta}$\\
     \hline
      0.2 & $1700 -14e^{4.2 \theta }$ & $7.9\times 10^7 e^{5.8 \theta }+2.1\times 10^{10}$ & $8.0\times 10^{10}-4.4\times 10^9 e^{3.1 \theta}$\\
     \hline
      0.1& $1700 -10 e^{4.5 \theta }$ & $9.7\times 10^7 e^{5.8 \theta }+2.5\times 10^{10}$ & $9.6\times 10^{10}-4.3\times 10^9 e^{3.3 \theta}$\\
     \hline
     all& $1-6.8\times 10^{-3} e^{4.4 \theta}$ & $5.2\times10^{-4}e^{7.4\theta}+1$ & $1.1 -0.055 e^{3.1 \theta}$\\
     \hline
      \bottomrule

\end{tabular}
\renewcommand{\arraystretch}{1}	
\caption{Angular fits for the peak Lorentz factor, gap half width, and outgoing photon energy flux. A representative selection of these fits are shown on Figs. \ref{normmassgap}, \ref{Bnormspec}, \ref{spinpropto}, and \ref{Ubpropto}.}
\label{tableall}
\end{table*}


\begin{thebibliography}{dum}
%[1]
\bibitem{1977MNRAS.179..433B}
R. D. Blandford and R. L. Znajek, Monthly Notices of the Royal Astronomical Society 179, 433 (1977).

\bibitem{Beskin:1992va} V. S. Beskin, Y. N. Istomin, and V. I. Parev, Soviet Astronomy (1992).

\bibitem{Hirotani:1998cf} K. Hirotani and I. Okamoto, The Astrophysical Journal 497, 563 (1998).

\bibitem{Levinson:2124471} A. Levinson and N. Globus, Monthly Notices of the Royal Astronomical Society 458 (2), 2269 (2016).

\bibitem{Globus:1604932} N. Globus and A. Levinson, Physical Review D 88, 084046 (2013).

\bibitem{Thorne:1986iy} K. S. Thorne, R. H. Price,  and D. A. Macdonald,
Black Holes: The Membrane Paradigm, (1986).

\bibitem{Macdonald:1984cd} D. A. Macdonald, Monthly Notices of the Royal Astronomical Society 211, 313 (1984).

\bibitem{Bardeen:1972fi} J. M. Bardeen, W. H. Press, and S. A. Teukolsky, The Astrophysical Journal 178, 347 (1972).

\bibitem{Prieto:2016eu} M. A. Prieto, J. A. Fern\'{a}ndez-Ontiveros, S. Markoff, D. Espada, and O. Gonz\'{a}lez-Mart\'{i}n, Monthly Notices of the Royal Astronomical Society 457, 3801 (2016).

\bibitem{0004-637X-736-2-103} L. W. Brenneman, C. S. Reynolds, and M. A. Nowak, The Astrophysical Journal 736, 103 (2011).

\bibitem{Genzel:1994iy}	R. Genzel, D. Hollenbach, and C. H. Townes, Reports on Progress in Physics 57, 417 (1994).

\bibitem{Dokuchaev:2016epe} V. I. Dokuchaev, Physics of Atomic Nuclei 78, 1517 (2015).

\bibitem{1538-4357-676-2-L109} J. M. Wang, Y. R. Li, and J. C. Wang, The Astrophysical Journal 676, L109 (2008).

\bibitem{0004-637X-786-1-5} M. Kino, F. Takahara, K. Hada, and A. Doi, The Astrophysical Journal 786, 5 (2014).

\bibitem{Johnson1242} M. D. Johnson, V. L. Fish, S. S. Doeleman, D. P. Marrone,  R. L. Plambeck, J. F. C. Wardle, K. Akiyama, K. Asada, C. Beaudoin, L. Blackburn, R. Blundell, G. C. Bower, C. Brinkerink, A. E. Broderick, R. Cappallo, A. A. Chael,  G. B. Crew, J. Dexter, M. Dexter, R. Freund, P. Friberg, R. Gold, M. A. Gurwell, P. T. P. Ho, M. Honma, M. Inoue, M. Kosowsky, T. P. Krichbaum, J. Lamb, A. Loeb, R. Lu, D. MacMahon, J. C. McKinney, J. M. Moran, R. Narayan,  R. A. Primiani, D. Psaltis, A. E. E. Rogers, K. Rosenfeld, J. SooHoo, R. P. J. Tilanus, M. Titus, L. Vertatschitsch, J. Weintroub, M. Wright, K. H. Young, J. A. Zensus, and L. M. Ziurys, Science 350, 1242 (2015).

\bibitem{Sikora07} Sikora M., Stawarz L., \& Lasota J. P., The Astrophysical Journal 658, 815 (2007).

\end{thebibliography}
\end{document}